\documentclass{cupconf}
\usepackage{natbib}
\usepackage{epsf}
\usepackage{psfig}
\usepackage{graphicx}
\usepackage{amsbsy}
\usepackage{amssymb}
\usepackage{amsmath}
\usepackage{bm}
\usepackage{eufrak}
\usepackage{upgreek}
\usepackage{revsymb}	
\usepackage{color}

\usepackage[OT2,T1]{fontenc}
\DeclareSymbolFont{cyrletters}{OT2}{wncyr}{m}{n}
\DeclareMathSymbol{\Sha}{\mathalpha}{cyrletters}{"58}

  \checkfont{eurm10}
  \iffontfound
    \IfFileExists{upmath.sty}
      {\typeout{^^JFound AMS Euler Roman fonts on the system,
                   using the 'upmath' package.^^J}%
       \usepackage{upmath}}
      {\typeout{^^JFound AMS Euler Roman fonts on the system, but you
                   dont seem to have the}%
       \typeout{'upmath' package installed. cupconf.cls can take advantage
                 of these fonts,^^Jif you use 'upmath' package.^^J}%
      }
  \else
  \fi

  \checkfont{msam10}
  \iffontfound
    \IfFileExists{amssymb.sty}
      {\typeout{^^JFound AMS Symbol fonts on the system, using the
                'amssymb' package.^^J}%
       \usepackage{amssymb}%
       \let\le=\leqslant  \let\leq=\leqslant
       \let\ge=\geqslant  \let\geq=\geqslant
      }{}
  \fi

  \IfFileExists{amsbsy.sty}
    {\typeout{^^JFound the 'amsbsy' package on the system, using it.^^J}%
     \usepackage{amsbsy}}
    {}

\newsavebox{\astrutbox}
\sbox{\astrutbox}{\rule[-5pt]{0pt}{20pt}}

\newtheorem{lemma}{Lemma}

\def\vec#1{\ensuremath{\mathchoice{\mbox{\boldmath$\displaystyle#1$}}
{\mbox{\boldmath$\textstyle#1$}}
{\mbox{\boldmath$\scriptstyle#1$}}
{\mbox{\boldmath$\scriptscriptstyle#1$}}}}

\def\sqr#1#2{{\vcenter{\vbox{\hrule height.#2pt
        \hbox{\vrule width.#2pt height#1pt \kern#1pt
                \vrule width.#2pt}
        \hrule height.#2pt}}}}

\title[Data analysis in asteroseismology]{A crash course on data analysis in asteroseismology\footnote{Lecture notes given at the 22$^{\rm th}$ Canary Islands Winter School, November 2010.}}

\author[T.Appourchaux]%
{T.Appourchaux}

\affiliation{Institut d'Astrophysique Spatiale\\
UMR8617, Universit\'e Paris-Sud\\
B\^atiment 121, 91405 Orsay Cedex, France}

\pubyear{2010}
\volume{838}
\pagerange{1--40}
\date{?? and in revised form ??}
\begin{document}
\maketitle

\begin{abstract}
In this course, I try to provide a few basics required for performing data analysis in asteroseismology.  First, I address how one can properly treat times series: the sampling, the filtering effect, the use of Fourier transform, the associated statistics.  Second, I address how one can apply statistics for decision making and for parameter estimation either in a frequentist of a Bayesian framework.  Last, I review how these basic principle have been applied (or {\it not}) in asteroseismology.

\end{abstract}

\begin{quotation}
\noindent Throughout human history, as our species has faced the frightening, terrorising fact that we do not know who we are, or where we are going in this ocean of chaos, it has been the authorities Ñ the political, the religious, the educational authorities who attempted to comfort us by giving us order, rules, regulations, informing -- {\it forming} in our minds -- their view of reality.   \\

\noindent To think for yourself you must question authority and learn how to put yourself in a state of vulnerable open-mindedness, chaotic, confused vulnerability to inform yourself.\\

Timothy Leary in {\it Sound Bites from the Counter Culture} (1989)
\end{quotation}

\section{Introduction}
This paper attempts to provide a summary of the course I gave during the 25$^{\rm th}$ Canaries Island Winter School.  In no way, this {\it course} should be perceived as the final answer to a problem.  I hope that this {\it course} can serve as a basis for students, fellow scientists to go beyond what is written here.  As in many approaches that I have pursued, this work is a snapshot of where I am and hopefully a possible starting point from which one can expand to other paths not yet ventured.

This {\it course} starts with a short historical introduction on signal processing and statistics or how our forefathers started doing data analysis more than 200 years ago.  The second part is related to the sampling and acquisition of continuous physical signals for subsequent analysis in a digital world.  The third part contains with a broad review of statistics from the so-called {\it frequentist} and {\it Bayesian} points of view.  The last part is related to the applications of the previous concept to data analysis for asteroseismology, which also includes a description of the physics behind that latter terms.

\section{Historical overview}
\label{sec:hist}
The basic principle of asteroseismic data analysis can be summarised as follows:
\begin{itemize}
	\item acquire signal from a finite world
	\item compute the Fourier transform of the discrete signal
	\item extract the characteristics of the harmonic signals
\end{itemize}
The first step is closely related to approximation of continuous function using decomposition on a base of orthogonal functions.  The latter could be the sine and cosine functions used in the second step, that is Fourier decomposition or transform.  The last step is related to inference based on statistics or applications of probability to the data.
Hereafter, I will try to place in a historical perspective all of these steps.  The perspective will be presented in a chronological order for each subject.  This historical review does not pretend to be complete as I am not an epistemologist.  The main goal of this historical review is to give the reader some keys on reflecting on some tools that we regularly use.  The reader will then be free to delve on the subject or leave it aside.

\subsection{Spectral analysis and Digital signal processing}
The father of spectral analysis is the well known Joseph Fourier.  He pioneered the decomposition of an arbitrary function in cosine and sine functions in his memoir on heat.  In Chapter 3 of his memoir (p. 257), \citet{Fourier} expressed the decomposition in harmonic functions that later became, using Euler's notation,  the complex Fourier transform.  In the original formulation of Fourier lies explicitly the potential for a finite summation.  In other words, any arbitrary function can be approximated with a finite summation over the Fourier coefficients, which is the Discrete Fourier Transform (DFT).  In essence, this is the introduction of a representation of a continuous world using a finite set: a {\it digital} world.  The expression provided by Fourier was already a digital description of the world.  The use of sine and cosine functions was very much at the center of the mathematical world at the beginning of the 19$^{\rm th}$ century.  In 1805, Carl F. Gauss devised an algorithm for interpolation of cosine and sine functions which would later be recognised as the Fast Fourier Transform (FFT), work which was posthumously published \citep{Gauss_fft}.  This work was published in Latin but translation can be found in \citet{Goldstine} and \citet{Heideman}.  The proposal 27 of Gauss' work is what we now call the FFT \citep{Heideman}.  The {\it original} algorithm was discovered (not to say re-discovered) by James Cooley and John Tuckey in 1965, while they were working at IBM and the Bell Telephone Laboratory, respectively \citep{CooleyTuckey}.  As anticipated by Gauss, they showed that the DFT could be speeded up by using the fact that the number of points $N$ in the transform could be expressed as a products of prime numbers, thereby speeding the computation by  $O\left({\frac{N}{\log N}}\right)$.  

At the time of the publication of the paper by \citet{CooleyTuckey}, spectral analysis had already entered the {\it modern} age of the digital world. When working at the AT\&T Bell Telephone Laboratory, Claude \citep{Shannon49} introduced the so-called sampling theorem which is key for reducing a continuous finite-bandwidth signal to a digital sample.  The frequency at which the signal should be sampled was derived by \citet{Nyquist24}, hence bearing the name the Nyquist frequency (Harry Nyquist belonged to what became later the Bell Telephone Laboratory).  The sampling theorem is at the basis of all digital audio equipment since the invention by Sony of the digital audio disc in 1972, later to become the Compact Disc of Philips in 1978.  Digital signal processing (DSP) is used in many applications such as speech recognition, compression, audio sampling, home cinema and of course in scientific processing.

It is rather surprising to realise that the existence of the digital world dates back to the beginning of the 19th century.  In a way, it is not so strange that we need to describe an infinite world with a finite set of data, for instance using function interpolation.  Being ourselves {\it finite} or limited in {\it time} and {\it space}, our world was bound to become sooner or later digital.
 
\subsection{Probability, statistics and inference}
Probability and statistics are related to one another.  Probability provides the mathematical foundations for assessing the chance that a random event will occur.  While statistics using probability theory provides inference on what has been {\it really} observed.  Probability theory started with Jakob Bernoulli who was applying combinatorial analysis for calculating probabilities related to the games of chance.  He published what is known as the Bernoulli distribution and also introduced the law of large numbers \citep{Bernoulli}.  The same Bernoulli distribution was approximated by \citet{DeMoivre} which was a special case of the Central Limit theorem.  Later on, Reverend Thomas Bayes solved a problem that was left untouched by \citet{DeMoivre} related to the probability of occurrence of unrelated events.  Proposition 5 of \citet{Bayes} is what is known today as the {\it Bayes theorem}.  This theorem was also found independently by \citet{Laplace} who was working on the same subject.  Unfortunately, this view on probability quite advanced at the times of Bayes and Laplace was not used until it was {\it re-discovered} by \citet{Jeffreys}.

Inference is related to how one can deduce from data a theoretical model of the world being observed (with error bars on this model).  The origin of the first inference can be traced back to the
work of Laplace, related to the use of the arithmetic mean \citep{Laplace}.   Gauss demonstrated, using the {\it Maximum Likelihood Principle}, that an estimate of a parameter measured many times can indeed be expressed as an arithmetic mean of these observations \citep{Gauss_mle}.   A typical inference called {\it Least Squares\footnote{translated literally from {\it Moindres quarr\'es}} minimisation} was used by \citet{Legendre} for deriving the orbit of comets, and for verifying the length of the meter through the measurement of the Earth's circumference.  This technique was also found before Legendre by Gauss but was published until later \citep{Gauss_mle}.  Gauss also derived what was called the law of errors: the so-called Gaussian distribution of errors \citep{Gauss_mle}.  The use of this ubiquitous error distribution is a simple consequence of the principle of the{\it Maximum Entropy Distribution}  \citep{Jaynes_B}; the distribution simply reflecting the state of our knowledge (or lack of) by knowing the mean value $\mu$ and the root means square deviation $\sigma$ of a set of observations.   The principles behind {\it maximisation} are at the very heart of inference.  While Gauss introduced the concept for likelihood, Fisher set the proper mathematical background and theory behind the use of {\it Maximum Likelihood Estimators}, notably their asymptotic properties and their information content \citep{Fisher, Fisher_info}.  Around that time, a controversy between Ronald Fisher and Harold Jeffreys marked the start of different view on probability and statistics: {\it frequentists} vs {\it Bayesian}.  \citet{Jeffreys} was key in reviving the approach touched upon by Bayes and Laplace, an approach that was not the main stream of statistical thinking at the time of Fisher.  At that point in time, the way was open for Bayesian probability and statistics to be applied in various fields of physics and astrophysics.  \citet{Jaynes_B} provides many possible applications of Bayesian approaches such as one used for Fourier analysis.

Now in the 21$^{\rm th}$ century, it would be naive to believe that inference can only be based on Bayesian approaches, it is surely not a {\it panacea}.  It should be borne in mind that as much as we evolve, any field of Science evolves accordingly.  Even in statistics the evolution is not finished.  The current stream is to try to reconcile the various approaches promoted by Fisher and Jeffreys \citep{Berger2003}.  Therefore, it is the responsibility of any physicist or astrophysicist to follow this evolution.  It is perhaps superfluous to remind the reader that all inferences on the world as we see it come from instrumentation, observation, phenomena that are neither perfect nor deterministic.  Let me remind you of a quote by Henri Poincar\'e: {\it From this point of view all the sciences would only be unconscious applications of the calculus of probabilities. And if this calculus be condemned, then the whole of the sciences must also be condemned} \citep{Poincare}.

\section{Digital signal processing and spectral analysis}

\subsection{Time series sampling}
\begin{figure}
\center{
 \hbox{ \includegraphics[width=0.35\textwidth,angle=90]{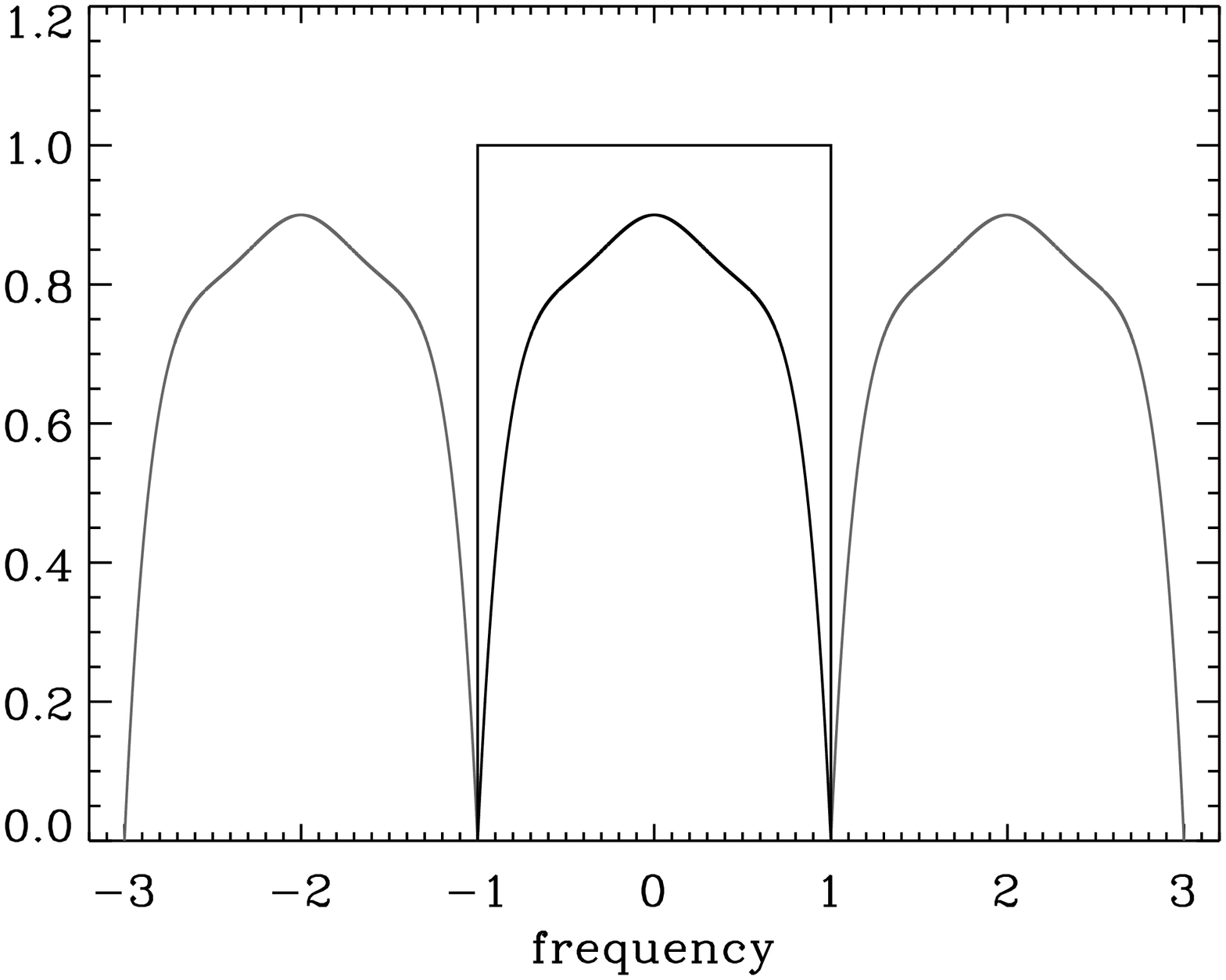}
\vspace{0.0truecm}
\includegraphics[width=0.35\textwidth,angle=90]{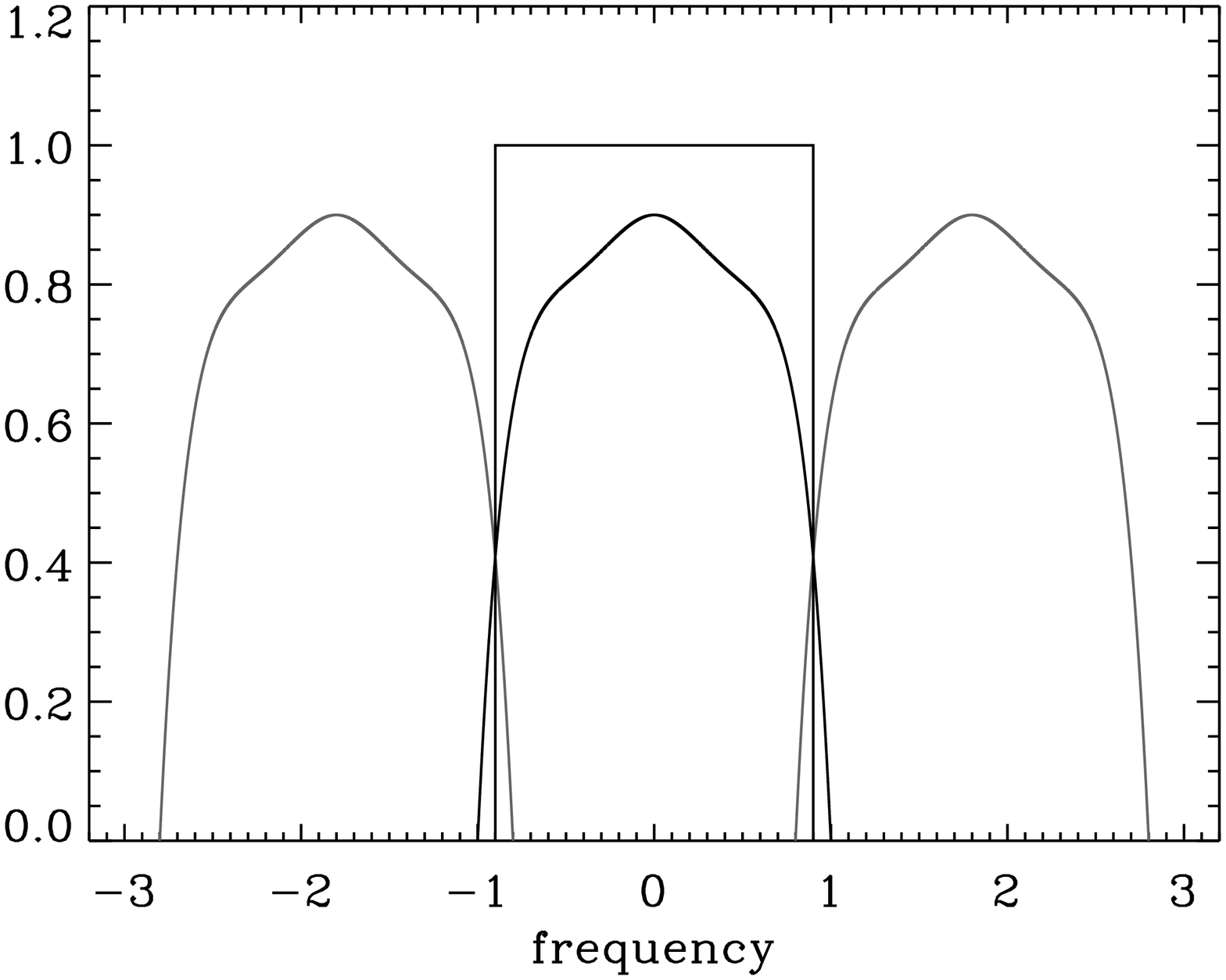}}
}

 \caption[]{The frequency response of a finite bandwidth signal with the boxcar on top, properly sampled according to Shannon theorem (Left), undersampled and aliased (Right)}

\label{shannon}
\end{figure}

\citet{Shannon49} provided the theorem which allows to sample a continuous signal whose frequencies are contained in a finite bandwidth.  Using the decomposition in Fourier
series, \citet{Shannon49} wrote that: {\it If a function $x(t)$ contains no frequencies higher than $\Delta \nu$, it is completely determined by giving
its ordinates at a series of points spaced 1/(2$\Delta \nu$) apart}", then he wrote:
\begin{equation}
x(t_{n})=\int_{-\Delta \nu}^{+\Delta \nu} X(\nu) {\rm e}^{{\rm i} 2 \pi \nu t_{n}} {\rm d} {\nu}
\end{equation}
where $t_{n}=n \Delta t$ (with $\Delta t=1/2\Delta \nu$), $x(t)$ is the function to be sampled, $X(\nu)$ is the Fourier transform of $x(t)$.  From this theorem, and the use of Fourier 
decomposition, one can then write:
\begin{equation}
X(\nu)=\Pi(2\Delta \nu) \left[\Delta t \sum_{n=-\infty}^{n=+\infty} x(t_n) {\rm e}^{{\rm i} 2 \pi \nu t_{n}}\right]
\label{Xnu}
\end{equation}
where $\Pi$ is the boxcar.  The term between brackets is simply the original Fourier decomposition which is implicitly periodic, hence the use of the
boxcar for delimiting the frequency space.  This is the case represented by the left hand side of Figure~\ref{shannon}.  Using the inverse Fourier transform of Eq.~(\ref{Xnu}), one can shows that we can fully reconstruct $x(t)$ by writing:
\begin{equation}
x(t)=\sum_{n=-\infty}^{n=+\infty} x(t_n) {\rm sinc} \left(\frac{t-t_n}{\Delta t}\right)
\end{equation}
where {\rm sinc} (=$\sin x /x$) is the sinus cardinal function.  This equation shows that one can recover {\it perfectly} a continuous function using samples of that function at regular spacing, whose cadence is provided by the spectral content of that function.  In practice, the recovery can only be approximated as the summation over $\infty$ is impractical.

\subsection{Aliasing}
There are two cases that provide unwanted frequency leaking into the original spectrum:
\begin{itemize}
\item Undersampling of a band-limited signal
\item Non band-limited signal
\end{itemize}
\noindent This is what is called {\it aliasing} whose effect is shown on the right hand side of Fig.~\ref{shannon}.    In this case high frequency signal leaks, back or aliases, at frequency $\nu_{\rm alias}$=$\nu_{\rm true}-\Delta \nu$.  Undersampling occurs when the sampling time is larger than Shannon's sampling time ($1/2 \Delta \nu$).  Undersampling of a band-limited signal can be easily resolved by applying Shannon's theorem.  The case of signal having non-limited frequency content is clearly not covered by Shannon's theorem.  In that case, there are two techniques that can provide a reduction of the aliasing power: integration and weights.  

\begin{figure}
\center{
 \hbox{ \includegraphics[width=0.35\textwidth,angle=90]{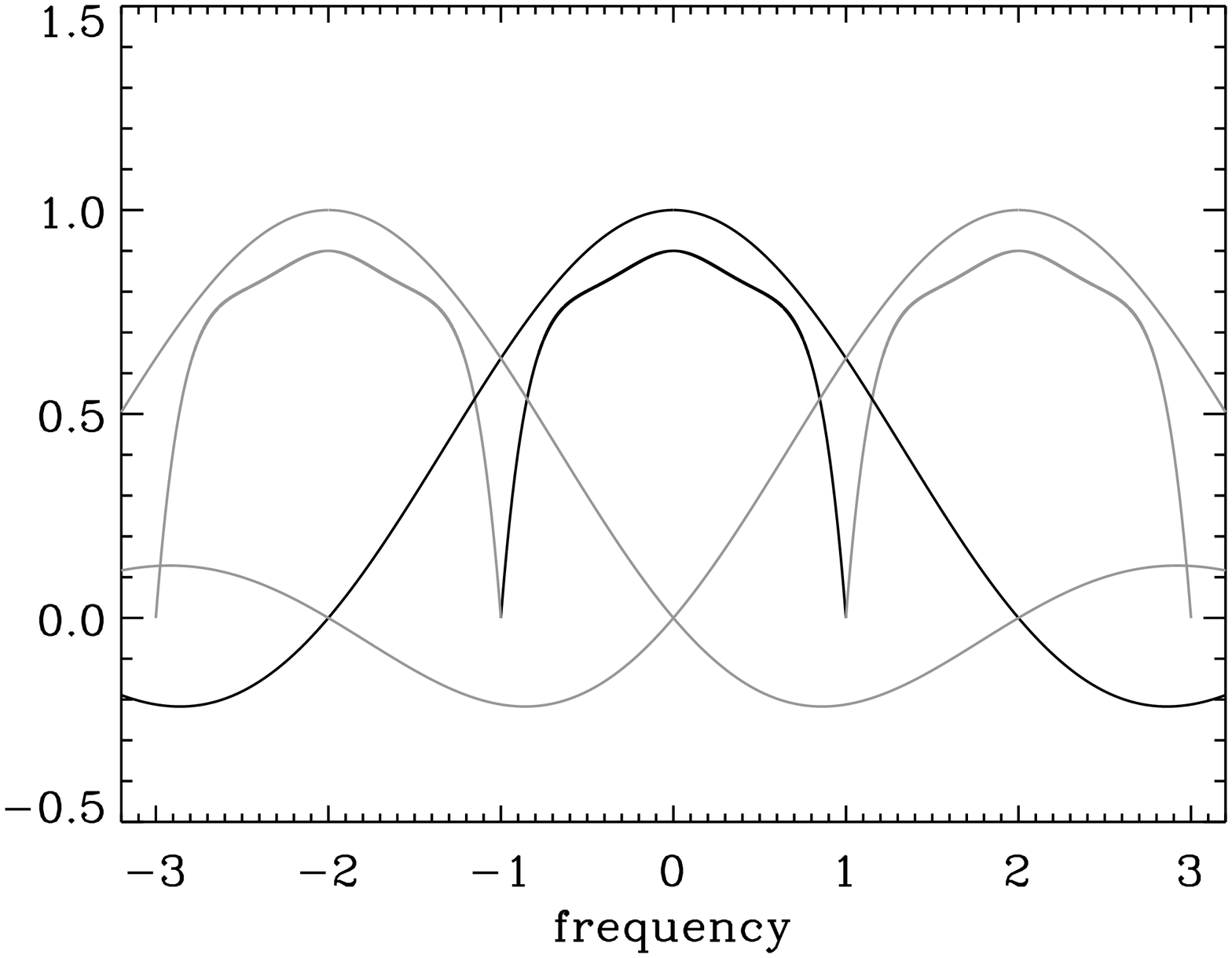}
\vspace{0.0truecm}
\includegraphics[width=0.35\textwidth,angle=90]{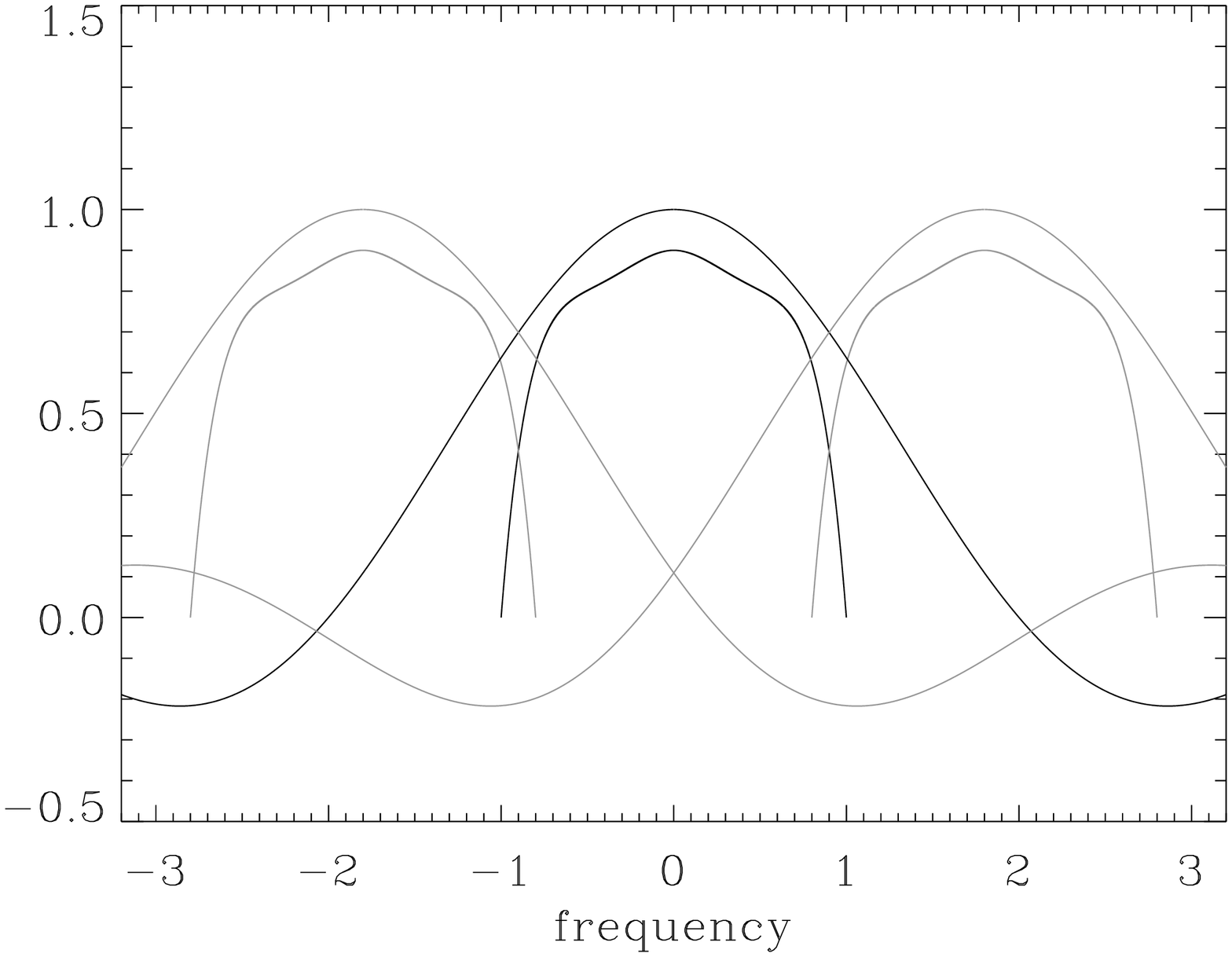}}
}
 \caption[]{The finite bandwidth signal and the sinc function related to the integration over 100\% of the sampling time (in black at $\nu=0$), (in grey at $\nu=\pm 1/2\Delta \nu$); when the signal is sampled according to Shannon's theorem (Left), when the signal is undersampled (Right).}

\label{integration}
\end{figure}

I shall focus on the effect of integration that is often encountered either when observing time series or making images with discrete arrays such as Charge Coupled Devices (CCD).  When one integrates a signal $x$ over some time ($\Delta t$) I can write:
\begin{equation}
x_{\rm obs}(t)= \sum_n \left[\frac{1}{\Delta t} \int_{t_n}^{t_n+\Delta t} x(t){\rm d}t \right]
\end{equation}
where $x$ is regularly sampled at $\Delta t_s=t_{n+1}-t_n$
This equation can be rewritten using convolution as:
\begin{equation}
x_{\rm obs}(t)=\Sha(\Delta t_s) * \left[\frac{1}{\Delta t}(x * \Pi(\Delta t))(t)\right]
\end{equation}
where $\Sha$ is the Dirac comb.  Since the Fourier transform of a Dirac comb is also a Dirac comb, the Fourier spectrum of the signal can then be written as:
\begin{equation}
X_{\rm obs}(\nu)=\frac{1}{\Delta t_s}{\Sha}\left(\frac{1}{\Delta t_s}\right) \left[X(\nu) {\rm sinc}(\Delta t \nu){\rm e}^{{\rm i}\pi \nu \Delta t}\right] 
\end{equation}
Figure~\ref{integration} shows the resulting effect of the integration when $\Delta t=\Delta t_s$.  In practice, when the signal is band limited,
integration will introduce a filtering effect of the high frequencies.  This effect can only be reduced by having a very short integration time with respect to the 
sampling or $\Delta t \ll \Delta t_s$.  The effect of aliasing is also shown on Fig.~\ref{integration}.  When the signal is non-band limited, there are two possible solutions
for reducing the effect of the high-frequency signal:
\begin{itemize}
\item Introduction of a window function in frequency (the $\Pi$ function)
\item Integration at 100\% duty cycle ($\Delta t=\Delta t_s$)
\end{itemize}
\noindent The first solution requires the combination of the time sample by using the Fourier transform of the $\Pi$ function which is a sinc in time.  This solution can only be
partially implemented as it would require a sum over $\infty$.  This solution is naturally implemented when making
images through a telescope using discrete arrays.  In that latter case, the highest spatial frequencies are cut off by having the telescope diameter providing a cut-off $D/\lambda$
half that of the spatial pixel sampling; this is illustrated by the left hand side of Fig.~\ref{integration}.  In other terms, there is no aliasing in a telescope when the pixel resolution
is half of the telescope resolution, that is two samples per resolution element.  De facto, in a telescope, the second solution is also used in combination with the first solution.  The second solution, although not perfect, will reduce 
the amplitude of the high frequency noise in time series.

\subsection{Filtering}
\begin{figure}
\center{
 \hbox{ \includegraphics[width=0.35\textwidth,angle=90]{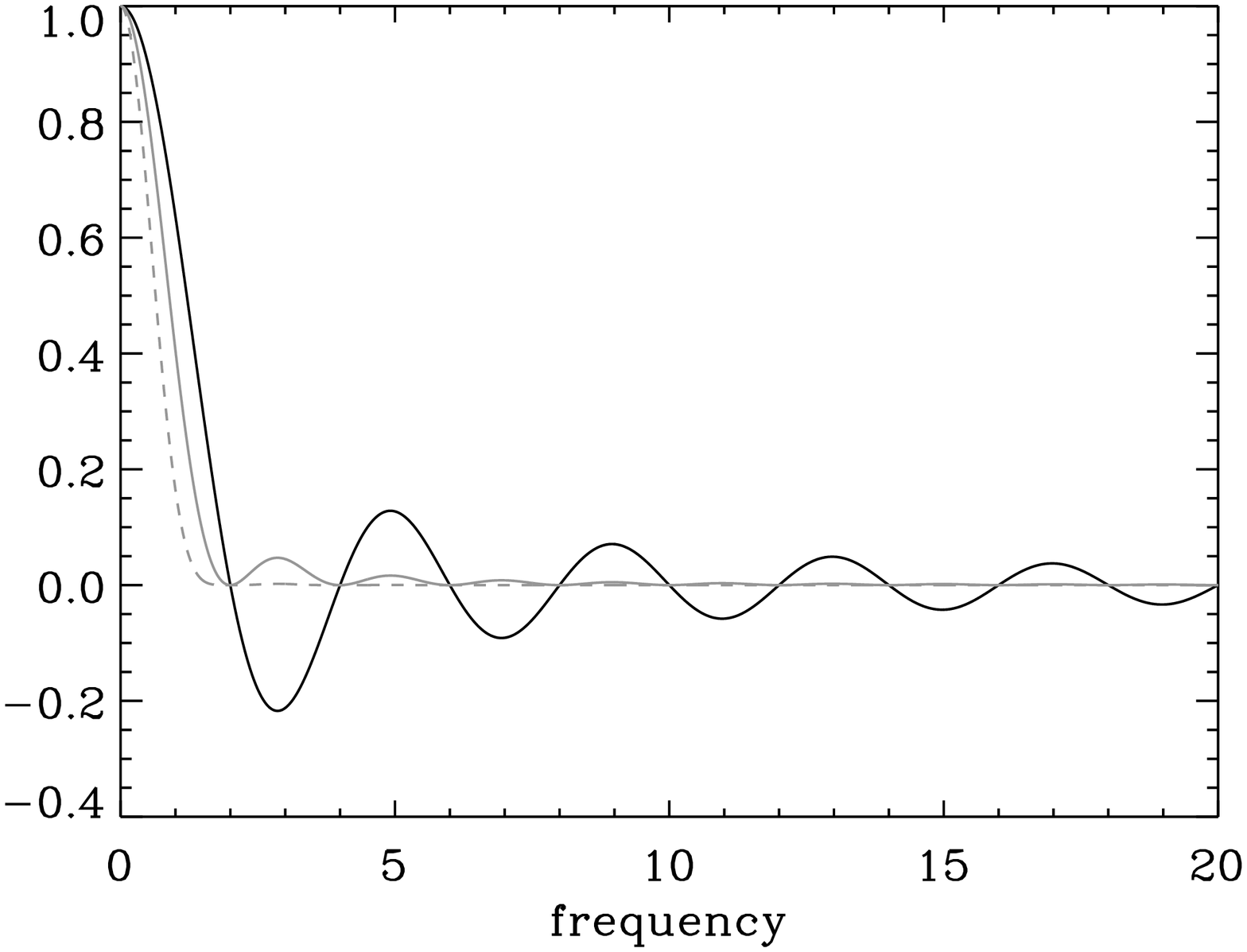}
\vspace{0.0truecm}
\includegraphics[width=0.35\textwidth,angle=90]{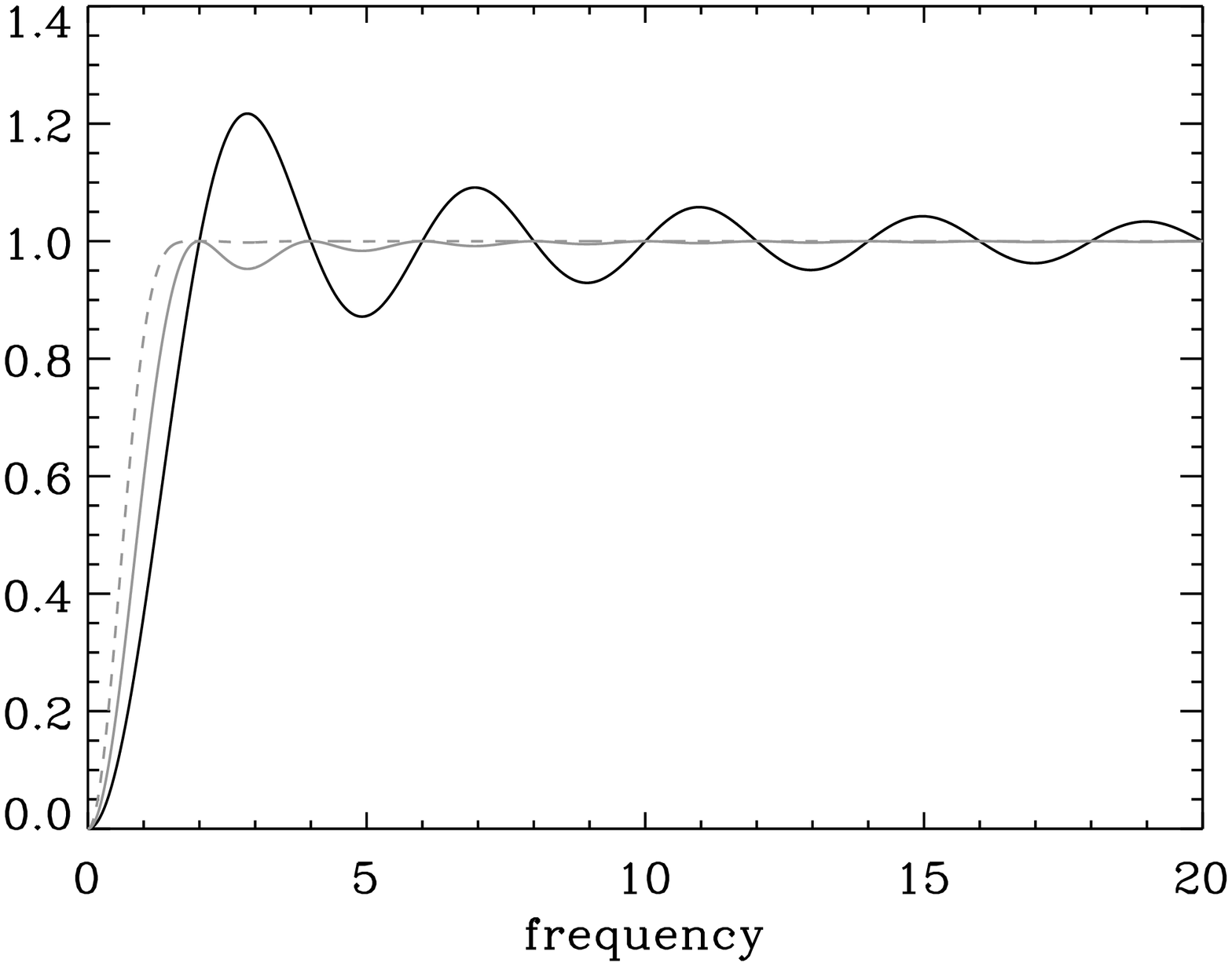}}
}
 \caption[]{(Left) Frequency response of several low pass filter: boxcar (black), triangle or 2 times the boxcar (grey), bell shape of 4 times the boxcar (dashed grey).  (Right) Frequency response of several high pass filter resulting from the previous low pass filters corresponding to the same color coding.}

\label{smooth}
\end{figure}

The effect of time series filters can simply be studied by understanding the effect of smoothing on time series.  Smoothing can be understand as
being the application of a weighting function sliding in time: a convolution.  This can be written as:
\begin{equation}
x_{\rm sm}(t)= (x * w) (t)
\end{equation}
where $w$ is the weighting function, which can be complex.  Using the Fourier transform, this becomes:
\begin{equation}
X_{\rm sm}(\nu)= X(\nu) W(\nu)
\end{equation}
where $W$ is the Fourier transform of $w$.   The smoothing filter typically provides a low pass filter which can be used to derive a high pass
filter of the original function $x$ by computing:
\begin{equation}
x_{\rm fil}(t)=x(t)-x_{\rm sm}(t)
\end{equation}
Using the Fourier transform, I get:
\begin{equation}
X_{\rm fil}(\nu)=X(\nu)(1-W(\nu))
\label{filter}
\end{equation}
Figure~\ref{smooth} shows the results of applying one time, two times and four times the boxcar on a time series.  Applying twice the boxcar
is equivalent to a triangular weighting function (convolution of 2 boxcar function), while applying four times the boxcar is equivalent to a bell shape weighting
function (convolution of 2 triangle functions).  It is rather clear from Fig.~\ref{smooth} that boxcar smoothing should be avoided because it provides too much
{\it ringing} effect of the type \citet{Gibbs98,Gibbs99} discovered.  The introduction of a less sharp transition for all derivatives by multiple boxcar smoothing provides a neat solution to this Gibbs effect.

Another form of filtering is to use the original series shifted in time by $t_0$ and then subtract the shifted series from the unshifted.  In that case the weighting function is simply the Dirac distribution $w(t)=\delta(t-t_0)$.  Then using Eq.~(\ref{filter}), the modulus of the filter is then simply
\begin{equation}
|1-W(\nu)|= 2 \sin (\pi \nu t_0)
\end{equation}
The frequency at which the transmission if half is given by $\nu_{\rm cuton}=1/3 t_0$.  This kind of smoothing filter has been used for the data of the Global Oscillation Network Group (GONG) using the first difference obtained with $t_0=\Delta t_s$ \citep{TA2000}.

\subsection{Time limits and the Discrete Fourier Transform}
\begin{figure}
\center{
\includegraphics[width=0.5\textwidth,angle=90]{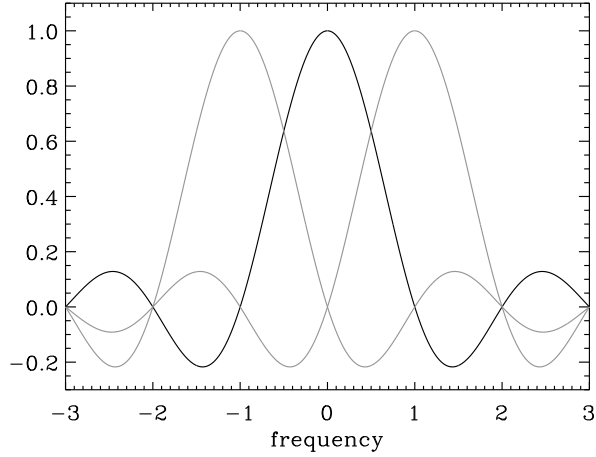}
}
 \caption[]{The sinc function as a function of frequency normalised to the resolution ($1/T$), for $\nu=0$ (black), for $\nu=\pm 1$ (grey)}
\label{sinc}
\end{figure}

The Fourier transform is not applicable when doing {\it real} data analysis.  There is no way that we can observe an infinite strings of data.  We usually observe during a finite time $T$ for which we
can compute the following Fourier transform:
\begin{equation}
X_{T}(\nu)=\int_{-T/2}^{+T/2} x(t) {\rm e}^{{\rm i} 2 \pi \nu t} {\rm d} t
\label{DFT1}
\end{equation}
which can be rewritten as:
\begin{equation}
X_{T}(\nu)= [X * {\rm sinc}(T)](\nu)
\end{equation}
The convolution function of the term in bracket is the sinc function.  Figure~\ref{sinc} shows the sinc function for adjacent frequency spaced at $\pm \frac{1}{T}$.  It is obvious
that the spectrum is correlated between the various frequency bins.  I will show later that the correlation is in fact null for frequency bins separated by integer values of $\frac{1}{T}$, and for slowly varying power spectra.  If I combine, the finite observation with a finite bandwidth signal, I then have the Discrete Fourier Transform (DFT):
\begin{equation}
X_{\rm DFT} \left(\nu_p\right)= \Delta t \sum_{n=1}^{n=N} x(t_n) {\rm e}^{{\rm i} 2 \pi \nu_p t_n}  
\label{DFT}
\end{equation}
with $\nu_p=\frac{p}{T}$, $t_n=n \Delta t_s$, $T=N \Delta t_s$.  Equation~(\ref{DFT}) is simply the truncated original Fourier series, which is then by definition periodic with period of $\Delta \nu=\frac{1}{t_s}$.  This property directly gives that $N$ is also the maximum value of $p$.  The DFT can be computed using the definition given above but it is rather time consuming as the time for computing the summation scales as $N^2$.  As mentioned in Section 2., \citet{CooleyTuckey} provided a faster way based on the factorisation properties of the Fourier transform; in that case the time for computing the summations scales as $N \log N$.

\subsection{Fourier transform of stationary processes}
The application of Fourier transforms to non deterministic or random functions is key in doing data analysis in astrophysics.  For such a process, I define for the random variable $x$, the {\it power spectral density} as:
\begin{equation}
S_T(\nu)=\frac{1}{T}|X_{T}(\nu)|^2
\end{equation}
where $X_T(\nu)$ is given by Eq.~(\ref{DFT1}).  I can then write $S_T(\nu)$ as:
\begin{equation}
S_T(\nu)=\frac{1}{T} \int_{-T/2}^{+T/2} \int_{-T/2}^{+T/2} x(t)x(t') {\rm e}^{{\rm i} 2 \pi \nu t} {\rm e}^{-{\rm i} 2 \pi \nu t'} {\rm d}t{\rm d}t'
\end{equation}
Since the process is stationary I can make the change of variable $\tau=t-t'$, and then I have:
\begin{equation}
S_T(\nu)=\int_{-T/2}^{+T/2} \left[\frac{1}{T} \int_{-T/2}^{+T/2} x(t')x(t'+\tau) {\rm d}t'\right]{\rm e}^{{\rm i} 2 \pi \nu \tau}  {\rm d}\tau
\end{equation}
The term in brackets is by definition the autocorrelation function ($C_T(\tau)$) of the process taken over a finite window $T$.  Then I have:
\begin{equation}
S_T(\nu)=\int_{-T/2}^{+T/2} C_T(\tau){\rm e}^{{\rm i} 2 \pi \nu \tau}  {\rm d}\tau
\label{Wiener-Khinchine}
\end{equation}
Equation~(\ref{Wiener-Khinchine}) introduces the so-called Wiener-Khinchin theorem which is essential for understanding the spectral analysis of random processes.  As a matter of fact, if 
I also assume ergodicity of the process (i.e. the temporal average can be exchanged with the spatial average) then I have:
\begin{equation}
C(\tau)={\rm E} (x(t)x(t+\tau))={\lim_{{T} \to \infty}} C_T(\tau)
\end{equation}
I can then write:
\begin{equation}
S_x(\nu)={\lim_{{T} \to \infty}} {\rm E}\left[S_T(\nu)\right]=\int_{-\infty}^{+\infty} C(\tau) {\rm e}^{{\rm i} 2 \pi \nu \tau}  {\rm d}\tau
\end{equation}
which is the proper Wiener-Khinchin theorem \citep{Wiener,Khinchin}.  Using the inverse Fourier transform, I can then also write:
\begin{equation}
C(\tau)=\int_{-\infty}^{+\infty} S_x(\nu) {\rm e}^{-{\rm i} 2 \pi \nu \tau}  {\rm d}\nu
\end{equation}
This relation between the auto-correlation function and the power spectral density is absolutely key in understanding the Fourier analysis of stationary processes.

\subsection{Parseval's theorem}
For $\tau=0$, I also have
\begin{equation}
C(0)=\int_{-\infty}^{+\infty} S_x(\nu) {\rm d}\nu={\rm E} (x^2)
\label{Parseval}
\end{equation}
This equation is also an alternate formulation of the energy conservation principle known as Parseval's theorem \citep{Parseval}.  This important property provided by Eq.~(\ref{Parseval}) can be adapted for normalising the DFT.  For the DFT given by Eq.~(\ref{DFT}), I can write:
\begin{equation}
\alpha^2 \sum_{p=1}^{p=N} |X_{\rm DFT} (\nu_p)|^2 = \frac{1}{N}\sum_{n=1}^{n=N} x(t_n)^2
\label{norm}
\end{equation}
where $\alpha$ is the required normalisation factor used for the spectral density.  For Eq.~(\ref{DFT}), it is easy to show that the normalization factor $\alpha$ is $1/N$ which is the usual factor used when computing the DFT.  If other definitions of the DFT are to be used, the proper normalisation factor $\alpha$ can be derived with Eq.~(\ref{norm}).  This latter equation is used for calibrating the spectra coming from different routines, all having different normalization factors.

\subsection{Statistics of the Discrete Fourier Transform}
\label{FFTstat}
Equation~(\ref{DFT}) is {\it our bread and butter} for extracting the frequencies associated with periodic phenomenon observed, for instance, in stars. 
Whenever one carries out observations, one should not forget that the observations come with noise either associated with the observed phenomenon or
with the instrument providing the data.  Therefore, it is essential to understand the statistics of the Fourier transform of random variables.  Let us start with a simple
and commonly used example: a random variable $x$ with an unknown distribution (with ${\rm E}(x)=0$ and ${\rm E}(x^2)=\sigma^2$) for which all time samples $x(t_n)$ are independent from each other, in addition the process is assumed to be stationary.  Using Eq.~(\ref{DFT}), I can then write the DFT of the $x(t_n)$ as:
\begin{equation}
X_{\rm DFT}^{\rm r} (\nu_p)+ i X_{\rm DFT}^{\rm i} (\nu_p)= \sum_{n=1}^{n=N} x(t_n) \cos(2 \pi \nu_p t_n)  + {\rm i}  \sum_{n=1}^{n=N} x(t_n) \sin(2 \pi \nu_p t_n)
\label{DFT2}
\end{equation}
where $X_{\rm DFT}^{\rm r}$ and $X_{\rm DFT}^{\rm i}$ are the real and imaginary part of the Fourier spectrum.  Since the $x(t_n)$ are independent and identically distributed (i.i.d.), by virtue
of the Central Limit Theorem, the statistical distribution of $X_{\rm DFT}^{\rm r}$ and $X_{\rm DFT}^{\rm i}$ is a normal distribution for $N \gg 1$ with:
\begin{equation}
{\rm E}(X_{\rm DFT}^{\rm r} (\nu_p))={\rm E}(X_{\rm DFT}^{\rm i} (\nu_p))=0  
\end{equation}
\begin{equation}
{\rm E}(\left[X_{\rm DFT}^{\rm r} (\nu_p)\right]^2)={\rm E}(\left[X_{\rm DFT}^{\rm i} (\nu_p)\right]^2)=\frac{N}{2} \sigma^2
\end{equation}
Then, since $X_{\rm DFT}^{\rm r}$ and $X_{\rm DFT}^{\rm i}$ are independent and have the same normal distribution, the statistics of the power spectrum is then by definition a $\chi^2$ with 2 degrees of freedom (d.o.f.).  

Unfortunately (or fortunately) none of the processes that we observe have the properties of being i.i.d.  The processes are usually stationary processes but not i.i.d.  because these processes have usually a memory such that the correlation of $x(t_n)$ and $x(t_m)$ are different from zero when $t_n \ne t_m$.  Nevertheless, in that case, it can be demonstrated that the components of the Fourier transform are {\it also} both normally distributed with the same mean of zero and the same variance, which depends upon frequency \citep{Peligrad}.  It is amusing to quote Peligrad and Wu on their finding: `{\it In this sense Theorem 2.1} [of \citet{Peligrad}] {\it justifies the folklore in the spectral domain analysis of time series: the Fourier transforms of stationary processes are asymptotically independent Gaussian.}''

\subsection{Time series sampled at unevenly times}
It is quite common in astrophysics to have samples that are not equally spaced in time.  This lack of uniformity could be due to a variable number of photon per seconds or due to variable detector read out time.  Usually, this can be avoided by carefully designing the electronics, see for instance \citet{CF97}.  Nevertheless, if the time series are unevenly sampled, there are ways and means to find solutions.  Equation~(\ref{DFT}) can still be applied, obviously by dropping $\Delta t_s$.   The problem of the latter equation is that for unevenly times the statistics of the Fourier spectrum is no longer $\chi^2$ with 2 d.o.f.  \citep{Scargle82}.  Equation~(\ref{DFT2}) can be adapted such that this statistical property is kept by writing:
\begin{equation}
X_{\rm LS}^{\rm r} \left(\nu_p\right)=  \frac{1}{w(\tau)} \sum_{n=1}^{n=N} x(t_n) \cos(2 \pi \nu_p (t_n-\tau))
\label{rLS}
\end{equation}
\begin{equation}
X_{\rm LS}^{\rm i} \left(\nu_p\right)=  \frac{1}{v(\tau)} \sum_{n=1}^{n=N} x(t_n) \sin(2 \pi \nu_p (t_n-\tau))
\label{iLS}
\end{equation}
where $w$ and $v$ are given by:
\begin{equation}
w(\tau)=  \sum_{n=1}^{n=N} \cos^2(2 \pi \nu_p (t_n-\tau))
\label{rLS_w}
\end{equation}
\begin{equation}
v(\tau)=  \sum_{n=1}^{n=N} \sin^2(2 \pi \nu_p (t_n-\tau))
\label{iLS_v}
\end{equation}
and $\tau$ is introduced for keeping the invariance in time of the transform given by Eqs.~(\ref{rLS}) and (\ref{iLS}).  $\tau$
is given by:
\begin{equation}
\tan(2 \pi \nu \tau)=\frac{\sum_{n=1}^{n=N} \sin (2 \pi \nu t_n)}{\sum_{n=1}^{n=N} \cos (2 \pi \nu t_n)}
\end{equation}
The definition provided by Eqs.~(\ref{rLS}) and (\ref{iLS}) has the benefit of giving a power spectrum or Lomb-Scargle (LS) periodogram ($\left[X_{\rm LS}^{\rm r} \left(\nu_p\right)\right]^2+\left[X_{\rm LS}^{\rm i} \left(\nu_p\right)\right]^2$) which is $\chi^2$ with 2 d.o.f. \citep{Scargle82}. It must also be pointed out that Eqs.~(\ref{rLS}) and (\ref{iLS}) are also the solution obtained when applying {\it Least Squares} minimisation to
\begin{equation}
\sum_{n=1}^{n=N} \left[x(t_n)-a_{\rm c} \cos(2 \pi \nu t)-a_{\rm s} \sin(2 \pi \nu t)\right]^2
\end{equation}
where $a_{\rm c}$ and $a_{\rm s}$ are given by Eqs.~(\ref{rLS}) and (\ref{iLS}), respectively.  For speed, the LS periodogram is usually computed using the implementation prescribed by
\citet{Press1989} which is an approximation of the LS periodogram based upon {\it extirpolation}\footnote{Reverse interpolation or extirpolation replaces a function value at any arbitrary point by
several function values on a regular mesh} on a regular mesh and the use of the FFT.  The prescription is then very close to interpolating
onto a regular mesh.  It is worth noting that most users of the LS periodogram for unevenly sampled data are in fact computing the
FFT of the original data resampled onto a regular mesh, but with a proper normalization as given by \citet{Scargle82}.  It must be noted that the LS periodogram does not provide a better solution to coping with the presence of gaps.  
The reason is that although the Fourier transform {\it explicitly} includes gaps as zeros, adding zeros is also {\it implicitly} performed with the LS periodogram.  As a consequence, correlations between frequency bins also exist with the LS periodogram, but these are generally ignored.  The correlations in the Fourier transform in the presence of gaps are addressed in the next section.

\subsection{The influence of gaps in the time series}
The impact of the gaps on the Fourier spectrum has been described by \citet{MG94}.  The gaps introduce correlation between frequency bins that need to be taken into account when one wants, for example, to fit the power spectrum \citep[See for applications][]{Stahn2008}.  Hereafter, I will provide the result regarding the correlation of the Fourier spectrum between the frequency bins.  Assuming that I observe, a random variable $x$ through a window $W$, the Fourier transform can be written as:
\begin{equation}
\tilde{\cal{X}}(\nu)=\int_{-\infty}^{+\infty} x(t) W(t) {\rm e}^{{\rm i} 2 \pi \nu t} {\rm d}t
\end{equation}
The mean correlation between two frequency bins $\nu_1$ and $\nu_2$ is given by:
\begin{equation}
{\rm E}[\tilde{\cal{X}}(\nu_1)\tilde{\cal{X}}^*(\nu_2)]={\rm E}\left[\int_{-\infty}^{+\infty}  \int_{-\infty}^{+\infty}  x(t) x(t') W(t)  W(t'){\rm e}^{{\rm i} 2 \pi \nu_1 t} {\rm e}^{-i 2 \pi \nu_2 t'} {\rm d}t {\rm d}t'  \right]
\label{aa}
\end{equation}
where * denotes the complex conjugate.  Following \citet{Gabriel1993}, I have the following properties for the real and imaginary parts of $\cal{X}$:
\begin{equation}
{\rm E}[\tilde{\cal{X}}_{\rm r}(\nu_1)\tilde{\cal{X}}_{\rm r}^*(\nu_2)]={\rm E}[\tilde{\cal{X}}_{\rm i}(\nu_1)\tilde{\cal{X}}_{\rm i}^*(\nu_2)]
\label{bb}
\end{equation}
\begin{equation}
{\rm E}[\tilde{\cal{X}}_{\rm r}(\nu_1)\tilde{\cal{X}}_{\rm i}^*(\nu_2)]=-{\rm E}[\tilde{\cal{X}}_{\rm i}(\nu_1)\tilde{\cal{X}}_{\rm r}^*(\nu_2)]
\label{cc}
\end{equation}
Using the Fourier transform of $x(t)$, I can rewrite Eq.~(\ref{aa}):
\begin{equation}
{\rm E}[\tilde{\cal{X}}(\nu_1)\tilde{\cal{X}}^*(\nu_2)]=\int \int \int \int  {\rm E}[X(\nu) X(\nu')] W(t) W(t'){\rm e}^{{\rm i} 2 \pi [(\nu_1-\nu') t - (\nu_2-\nu) t']} {\rm d}t {\rm d}t'  {\rm d}\nu {\rm d}\nu'
\end{equation}
By construction, I assume that there is no correlation between the real and imaginary parts of the original spectrum $X$, that their variances have the same value E$(X_{\rm r}^2(\nu)]$ and the frequency
bins of the original spectrum are not correlated \citep[See also,][]{Gabriel1993}, such that I have:
\begin{equation}
{\rm E}[X_{\rm r}(\nu)X_{\rm r}(\nu')]={\rm E}[X_{\rm r}^2(\nu)] \delta(\nu'-\nu)
\end{equation}
\begin{equation}
{\rm E}[X_{\rm r}(\nu)X_{\rm r}(\nu')]={\rm E}[X_{\rm i}(\nu)X_{\rm i}(\nu')]
\end{equation}
where $\delta$ is the Dirac distribution.  Then I can rewrite:
\begin{equation}
{\rm E}[\tilde{\cal{X}}(\nu_1)\tilde{\cal{X}}^*(\nu_2)]=2\int_{-\infty}^{+\infty}   {\rm E}[X_{\rm r}^2(\nu)] [W(\nu_1-\nu) W(\nu-\nu_2)] {\rm d}\nu
\label{corr}
\end{equation}
where $W$ is the Fourier transform of the window function $w$.  For understanding the impact of Eq.~(\ref{corr}), let us assume that we observe white noise of mean 0 and of variance $\sigma_0$ in frequency.  In that case, I have:
\begin{equation}
{\rm E}[\tilde{\cal{X}}(\nu_1)\tilde{\cal{X}}^*(\nu_2)]=2\sigma_0^2\int_{-\infty}^{+\infty}   W(\nu_1-\nu) W(\nu-\nu_2) {\rm d}\nu
\end{equation}
Using the properties of convolution and the inverse Fourier transform, it can be shown\footnote{I leave the demonstration to the reader} that I have:
\begin{equation}
{\rm E}[\tilde{\cal{X}}(\nu_1)\tilde{\cal{X}}^*(\nu_2)]=2\sigma_0^2\int_{-\infty}^{+\infty}   w^2(t) {\rm e}^{{\rm i} 2 \pi (\nu_1-\nu_2)t} {\rm d}\nu
\end{equation}
The integral in this equation is simply the Fourier transform of the square of the window function, or $W_{\rm sq}$.  For a window function such as the one provided by an observing window of length $T$ (See Eq.~\ref{DFT1}), the correlation is then given by the sinc function.  This justifies {\it a posteriori} the sampling at frequency interval of $1/T$, for which the correlation is null.  If the variations of ${\rm E}[X_{\rm r}^2(\nu)]$ are slow with respect to $W(\nu)$, I can also rewrite Eq.~(\ref{corr}) as:
\begin{equation}
{\rm E}[\tilde{\cal{X}}(\nu_1)\tilde{\cal{X}}^*(\nu_2)] \approx 2 {\rm E}[X_{\rm r}^2(\nu_1)] W_{\rm sq}(\nu_1-\nu_2)
\label{approx}
\end{equation}
where $W_{\rm sq}$ is the Fourier transform of $w^2$.  Of course this approximation does not hold when the variations in frequency are over scales of $1/T$.  Nevertheless, Eq.~(\ref{approx}) gives an interesting solution for understanding the correlation between frequency bins in a Fourier spectrum.  

It is also useful to understand the correlation between the real and imaginary parts of the Fourier transform.  Using Eqs.~(\ref{bb}) and (\ref{cc}), I can derive the very useful formulation:
\begin{equation}
{\rm E}[\tilde{\cal{X}}_{\rm r}(\nu_1)\tilde{\cal{X}}_{\rm r}^*(\nu_2)]=\int_{-\infty}^{+\infty}   {\rm E}[X_{\rm r}^2(\nu)] {\cal R}\left[W(\nu_1-\nu) W(\nu-\nu_2)\right] {\rm d}\nu
\label{corr1}
\end{equation}
\begin{equation}
{\rm E}[\tilde{\cal{X}}_{\rm r}(\nu_1)\tilde{\cal{X}}_{\rm i}^*(\nu_2)]=\int_{-\infty}^{+\infty}   {\rm E}[X_{\rm r}^2(\nu)] {\cal I}\left[W(\nu_1-\nu) W(\nu-\nu_2)\right] {\rm d}\nu
\label{corr2}
\end{equation}
where ${\cal R}$ and ${\cal I}$ denote the real and imaginary operators, respectively.  These two relations can be used when a specific window which is different from the boxcar
is used.  For example, the use of weights across the observing window will then introduce correlation provided by Eqs.(\ref{corr1}) and (\ref{corr2}).  The introduction of these weights
or tapers is described in the next section.

%The impact of the gaps for the average will have the same origin as for the Fourier spectra.  As for the smoothed spectra, there is an intrinsic correlation between the $q'$ bins; the number of truly %independent frequency bins is then divided by $q'$.  In that latter case, the gaps do not have a large influence on the smoothed spectrum unless the fraction of gaps is about $1/q'$.  For multitapered %spectra, the influence of gaps can be taken into account for obtaining optimised tapers that match the structure of the gaps \citep{Fodor1998}.  In that case, the correlation between the frequency bins, %although reduced, will be not negligible, especially if several tapers are used for the estimation of the mean power spectrum.

\subsection{Taper estimates of Fourier power spectrum}
Fourier spectrum estimation is well adapted for periodic signals (pure sine waves or 
stochastic waves) but not necessarily well suited for estimating the spectral
density of frequency-dependent noise (pink or red noise).  For that purpose, one can: 
 \begin{itemize}
 \item average the power spectrum over an ensemble of $n$ sub-series,
 \item smooth the power spectra over $n$  frequency bins or,
 \item use multitapered spectra using the full time series for deriving a similar average.  
 \end{itemize}

Fourier spectrum estimation can be replaced by multitapered
spectra that are widely used in geophysics \citep[for a review
see][]{thomson82}.  Multitapered spectra are generated by applying a
set of tapers to a single time series, and an estimate
of the mean power spectrum is derived from an average of these
spectra.  Using tapers due to \citet{Slepian1978},  the multitapered spectra are statistically independent from
one another, and the statistics of the mean spectrum follows a
$\chi^{2}$ distribution with 2$n$ degrees of freedom \citep[where $n$
is the number of tapers,][]{thomson82}.  While the statistics of the
average power spectrum (or smoothed power spectrum) also follow a
$\chi^{2}$ distribution with 2$n$ degrees of freedom: the resolution
of the average spectrum is $n$ times lower than that of the
multitapered spectrum.  In helioseismology the use of these slepian
tapers has been replaced by more practical (but less accurate) sine
tapers \citep{Komm99}.  Unfortunately, for sine waves, tapers
tend to broaden the peaks, as shown by \citet{thomson82}.  Tapers as
such provide more benefit for broader peaks than for narrower peaks.

\section{Data analysis and statistics}
\subsection{Hypothesis testing}
Statistical testing is essential when one wants to decide: {\it have
we found a signal} or {\it not}?  This is related to {\it decision theory},
which can be summarised as {\it how do we choose between one
hypothesis versus another in the presence of uncertainties?}  In this
area, there are two schools of thought: the frequentist school and the
Bayesian school.

The difference between a Bayesian and a frequentist relates to their views of {\it subjective} versus {\it objective} probabilities.  A
frequentist thinks that the laws of physics are {\it deterministic},
while a Bayesian ascribes a belief that the laws of physics are true
or {\it operational}.  The {\it subjective} approach to probability
was first coined by \citet{Finetti}.  

For the rest of us, the
difference in views between frequentists and Bayesians can be outlined by taking an example from the Six-nation rugby tournament.  For a frequentist,
France has been winning over England in their direct confrontation only 39\% of these matches since 1906.  Based on this result, for a frequentist, France
has only 39\% chance of winning any future game against England.  For a Bayesian, this is a complete different story.  A Bayesian may attribute higher
chance or lower chance to France to win a given game based on the current physical and technical skills of each player, on the current ability of the players to play as a team and
on the psychological and mental health of the players and of the team as a whole.  Based on this assessment, a Bayesian might have attributed 70\% chance to win the 2010 game, or 20\% chance to win the 2011 game.  This is an {\it a posteriori} evaluation of the chances as both games took place while writing this course.  It is used as an example.  

In short, frequentists assign
probability to measurable events that can be measured an infinite
number of times, while Bayesians assign probability to events that
cannot be measured, like the survival time of the human race \citep{Gott94} or like the future outcome of sporting matches.

In what follows, I will try to give an
overview of what {\it I believe} I know on a subject that is rapidly evolving; and what I write is certainly {\it not} gospel.

\subsubsection{Frequentist hypothesis testing}
\label{sec:hypothesis}
For a frequentist, statistical testing is related to hypothesis testing.  In short, we have two types of hypotheses:
 \begin{itemize}
 \item ${\rm H}_0$ hypothesis or null hypothesis: what has been observed is pure noise
 \item ${\rm H}_1$ hypothesis or alternative hypothesis: what has been observed is a signal
 \end{itemize}
For the ${\rm H}_0$ hypothesis, I assume known statistics for the random variable $Y$ observed as $y$ and 
assumed to be pure noise; and then set a {\it false alarm probability}
that defines the acceptance or rejection of the hypothesis.  The so-called {\it detection significance} (or
{\it p-value}, terms not widely used in astrophysics) is the
probability of having a value as extreme as {\it the one actually
observed}.  There is an on-going confusion because statisticians call
{\it the significance level} what astronomers call the {\it false
alarm probability}; and statisticians call the {\it p-value} what is set in
astronomy as the {\it detection significance} (which is {\it not} the
significance level).  Here I shall use the current vocabulary
understood in astronomy. For example, the {\it false alarm
probability} $p$ for the ${\rm H}_0$ hypothesis is defined as:
 \begin{equation}
 	p = P_0(\tilde{T}(Y) \geq \tilde{T}(y_{\rm c})),
 \end{equation}
where $\tilde{T}$ is the statistical test, and $P_0$ is the probability of
having $\tilde{T}(Y) \geq \tilde{T}(y_{\rm c})$ when ${\rm H}_0$ is true; and $y_{\rm c}$ is the cut-off threshold derived from the test $\tilde{T}$ and the value $p$.  For example, take
the case of a random variable $Y$ distributed with $\chi^2$, 2 degrees
of freedom (d.o.f) statistics, having a mean of $\sigma$. If I
further assume that $\tilde{T}(Y)=Y$, I then have that:
 \begin{equation}
	p = P_0(Y \geq y_{\rm c}) = {\rm e}^{-\frac{y_{\rm c}}{\sigma}}
\label{chi2H0}	
 \end{equation}
If one observes a value $\tilde{y}$ of the random variable $Y$ that is larger than $y_{\rm c}$, the ${\rm H}_0$ hypothesis is rejected.  The
value that is quoted in this case is the {\it detection significance}
${\cal D}$, i.e.,
 \begin{equation}
	{\cal D} = {\rm e}^{-\frac{\tilde{y}}{\sigma}}
	\label{p-value}
 \end{equation}
The ${\rm H}_0$ hypothesis was used by \citet{Scargle82} for setting a false alarm probability, and by \citet{TA2000} to impose an
upper limit on g-mode amplitudes.  The method was based on the
knowledge of the statistical distribution of the power spectrum of
full-disc asteroseimic instruments, namely the $\chi^2$
distribution with 2 d.o.f.  For the ${\rm H}_1$ hypothesis, I assume given statistics both for
the noise and for the signal that we wish to detect, and set a level
that defines the acceptance or rejection of that hypothesis.  

In this example, I took as given that the test $\tilde{T}$ was known ($\tilde{T}(Y)=Y$).  As a matter of fact, such a test is not obtained in an
{\it ad hoc} manner but can be rationally derived using the Neyman-Pearson lemma.  An example of the application of this lemma for deriving the test and level provided by Eq.~(\ref{chi2H0}) is explained in the next section.

\begin{figure}
\center{
\includegraphics[width=0.5\textwidth,angle=90]{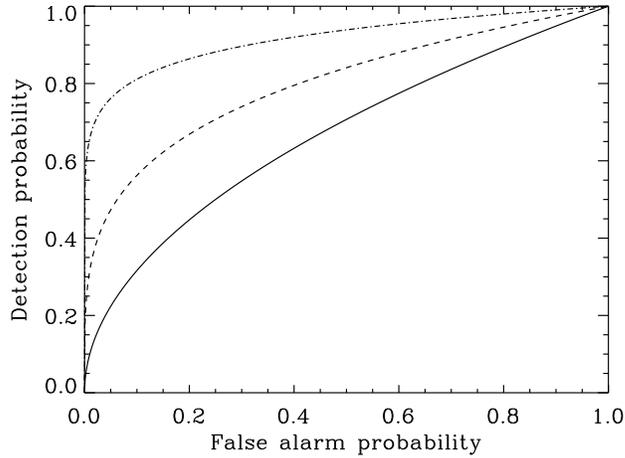}
}
 \caption[]{Detection probability as a function of the false alarm probability for pure sine waves stochastically excited for various signal-to-noise ratio: 1 (continuous line), 2 (dashed line), 10 (dot-dashed line)}
\label{roc}
\end{figure}

\noindent{\bf The Neyman-Pearson lemma.} The derivation of the best test and of the levels associated with the ${\rm H}_0$ and ${\rm H}_1$ hypotheses is provided by the Neyman-Pearson lemma \citep{NeymanPearson}.
This lemma is very useful for designing tests that will maximise signal detection while minimising noise effects.
\begin{lemma}
$\exists\,\eta>0$ {\rm such that} $\Lambda(y)=\frac{L(y | {\rm H}_0)}{L(y | {\rm H}_1)} \le \eta$ {\rm where} $P(\Lambda(y) \le \eta | {\rm H}_0)=\alpha$
\end{lemma}
\noindent where $L(y | {\rm H}_0)$ and $L(y | {\rm H}_1)$ are the likelihood for each hypothesis and $\alpha$ is called the {\it power} of the test.  I will show later how
one can use such a lemma for a specific  case that is often encountered in astronomy: the detection of single frequency peak in a power spectrum of a star having eigenmodes with very long lifetimes.   Let us assume that we observe
pure noise in a power spectrum,  since the statistics is $\chi^2$ with 2 d.o.f. I can write the likelihood of observing a value $\tilde{y}$ as:
 \begin{equation}
 L(\tilde{y} | {\rm H}_0)=\frac{1}{B}{\rm e}^{-\tilde{y}/B}
 \end{equation}
where $B$ is the mean noise level in the power spectrum.  Next, I assume that the peak is not deterministic but that its amplitude
is stochastic with an amplitude $A$. The likelihood for ${\rm H}_1$ is then:
 \begin{equation}
 L(\tilde{y} | {\rm H}_1)=\frac{1}{B+A}{\rm e}^{-\tilde{y}/(B+A)}
 \end{equation}
The likelihood ratio then can be written as:
\begin{equation}
\Lambda(\tilde{y})=(1+H){\rm e}^{-\frac{\tilde{y} H}{1+H}}
\label{likelihood}
\end{equation}
with $H=A/B$.  Then applying the Neyman-Pearson lemma leads to:
\begin{equation}
\Lambda(\tilde{y}) \le \eta \Rightarrow \tilde{y} \geqslant \eta'
\end{equation}
where $\eta'$ is given by solving
\begin{equation}
P(\tilde{y} \geqslant \eta' | {\rm H}_0)={\rm e}^{-\frac{\eta'}{B}}=\alpha
\end{equation}
which justifies {\it a posteriori} the use of Eq.~(\ref{chi2H0}).  Then I can also write the detection probability of a sine waves as:
\begin{equation}
P(\tilde{y} \geqslant \eta' | {\rm H}_1)={\rm e}^{-\frac{\eta'}{A+B}}=\alpha^{\frac{1}{1+H}}
\end{equation}
Figure~\ref{roc} shows the result for the detection probability for sine waves stochastically excited.  Such a diagram is also called the {\it receiver operating characteristic} (roc).  It provides a very efficient way of assessing the performance of the statistical test used.  In summary, the Neyman-Pearson lemma can be used for deriving in a non-arbitrary fashion the best test for accepting/rejecting ${\rm H}_0$.  With this lemma, the design of a test is therefore more systematic and less prone to improvisation.

\begin{table}
\caption{Types of error obtained for different decisions, based upon
the statistical test performed, and how the error relates to the
status of the ${\rm H}_0$ hypothesis.}
\centering
\begin{tabular}{cccc}
\hline\noalign{\smallskip}
%first & second & third  \\[3pt]
%\tableheadseprule\noalign{\smallskip}
&&\multicolumn{2}{c}{{\bf Status of H$_0$}}\\[1ex]
&&{\it True}&{\it False}\\[1ex]
\hline\noalign{\smallskip}
&{\it Reject}& Type I& Correct\\
\raisebox{1.5ex}{{\bf Decision}}&{\it Accept}&Correct&Type II\\
\noalign{\smallskip}\hline
\end{tabular}    % Give a unique label
\label{H_0}
\end{table}

\noindent{\bf Is the world dichotomic?} Taking a decision based on the result given by a single test, for
either hypothesis, could lead to errors in the decision process.  For
instance, the null hypothesis could be wrongly rejected when it is
true ({\it false positive} or wrong detection), but could also be
wrongly accepted while it is false ({\it false negative} or no
detection in presence of a signal).  The {\it false positive} results
in a {\it Type I} error, while {\it the false negative} results in a
{\it Type II} error (see Table~\ref{H_0}).  The ideal case would be, using the Neyman-Pearson lemma, to set a
test that would minimise the occurrence of both types of errors

It has been customary when applying the ${\rm H}_0$ hypothesis to set the decision level arbitrarily at 10\%
\citep{TA2000}.  From the frequentist view point there is nothing wrong in setting {\it a priori} the decision level before the test is
applied.  There are three types of result we might obtain from applying the test:
 \begin{enumerate}
 \item ${\rm H}_0$ always rejected
 \item ${\rm H}_0$ rejected or accepted at a level very {\it close} to 10\%
 \item ${\rm H}_0$ always accepted
 \end{enumerate}
Decision (a) will lead to the mention of a {\it detection being
statistically significant} at a level provided by the {\it detection
significance} (for example from Eq.~(\ref{p-value})).  The question is then to know what was detected. 
 The next step would then be the application of a test for the ${\rm H}_1$ hypothesis taking into account assumptions about
the detected signal, which may very likely result in the detection of signal.   Decision (c)
seems straightforward, i.e., noise dominates, but might one then be
tempted to lower, {\it a posteriori}, the decision level?  Decision (b) is the more
difficult borderline case, forcing us to either accept or reject ${\rm
H}_0$.  Here, we might ask: are things really that clear cut?  What
are the chances that if we accept ${\rm H}_0$ it is actually wrong
(Type II error), or truly right if rejected (Type I error)?  

These potential actions result from the application of a frequentist
test trying to answer the following question: what is the likelihood
of the observed data set $\tilde{y}$, given that ${\rm H}_0$ is true
or $p(\tilde{y} | {\rm H}_0)$?  The {\it detection significance}
mentioned when the test rejects the ${\rm H}_0$ hypothesis is nothing
but $p(\tilde{y} | {\rm H}_0)$, when actually what we want to know is
the likelihood that ${\rm H}_0$ is true given the data, i.e., $p({\rm
H}_0 | \tilde{y})$ ($\neq p(\tilde{y} | {\rm H}_0)$).  The frequentist view does
provide a useful answer when one can repeat the observations ad
infinitum.  But when we have only one universe, one observation, another approach must be used based upon Bayes' theorem; an approach which in principle gives access
directly to $p({\rm H}_0 | \tilde{y})$.

\subsubsection{Bayesian hypothesis testing}

\noindent{\bf On the posterior probability.} We should never forget the {\it two
sides of the coin}: if probability (likelihood) can justify {\it
alone} the rejection or acceptance of an hypothesis, this probability
{\it is not} the significance that the hypothesis is rejected or accepted.
The decision levels discussed above are related directly to a
well-known controversy in the medical field, concerning improper use
of Fisher's p-values as measures of the probability of effectiveness
of a medicine or drug \citep{Sellke2001}.  The {\it detection
significance} (or p-value) is improperly used as the significance of
the evidence against the null hypothesis.  It is far from trivial at
first sight to understand what is wrong with the {\it detection
significance}. Let us recall the example I gave above for a random variable $Y$ having a $\chi^2$ with 2 d.o.f statistics.  In that
case the {\it detection significance} is given as:
 \begin{equation}
	{\cal D} = {\rm e}^{-\frac{\tilde{y}}{\sigma}} \not\equiv P_0(Y \geq \tilde{y}).
	\label{p-value1}
 \end{equation}
 The latter statement ($\not\equiv$) is fundamental.  The observation is performed only once providing a value of $\tilde{y}$ and hence the detection significance ${\cal D}$.  But in no way does
 it provide the probability that the random variable is {\it always} above $\tilde{y}$ (or $P_0(Y \geq \tilde{y})$).
It is not correct to assume that if the observation were repeated it would provide the same level $\tilde{y}$.
 The mistake is to ascribe a significance to a
measurement performed only once, i.e., not repeated, and spanning just
a very small volume of the parameter space (e.g. $Y \in
[\tilde{y},\tilde{y}+\delta y]$).  If one makes a measurement $\tilde{y}$ of the random
variable $Y$ that is above $y_{\rm c}$, the significance of that measurement
is {\it not} ${\rm e}^{-\tilde{y}/\sigma}$.  In the framework of Bayesian
statistics, we are not interested in the {\it detection significance}
but in the posterior probability of the hypothesis $p({\rm H}_0 | \tilde{y})$, in other words as
already stated above $p({\rm H}_0 | \tilde{y}) \neq p(\tilde{y} | {\rm H}_0)$.  A similar description of this misunderstanding has been presented by \citet{Sturrock2009}.

In order to derive the posterior probability $p({\rm H}_0 | y)$, let
us first recall the Bayes' theorem.  The theorem of \citet{Bayes} relates the probability of an event A given the occurrence of an event B to the probability of the event B given the occurrence of the event A, and the probability of occurrence of the events A and B alone.
\begin{equation}
P({\rm A} | {\rm B}) = \frac{P({\rm B} | {\rm A}) P({\rm A})}{P({\rm B})}
\label{Bayes}
\end{equation}
For example, the probability of having rain given the presence of clouds is related to the probability of having clouds given the presence of rain by Eq~(\ref{Bayes}).  The term {\it prior probability} is given to $P({\rm A})$ (probability of having rain in general).  The term {\it likelihood} is given to $P({\rm B} | {\rm A})$ (probability of having clouds given the presence of rain) .  The term {\it posterior probability} is given to $P({\rm A} | {\rm B})$ (probability of having rain given the presence of clouds).  The term {\it normalization constant} is given to $P({\rm B})$ (probability of having clouds in general).  

The posterior probability of a
hypothesis H, given the data D and all other prior information I, is stated as:
 \begin{equation}\label{bayes}
    P({\rm H} | {\rm D,I}) = \frac{P({\rm H} | {\rm I})P({\rm D} | {\rm H,I})}{P({\rm D} | {\rm I})}.
 \end{equation}
where $P({\rm H} | {\rm I})$ is the prior probability of H given I, or
otherwise known as the prior; $P({\rm D} | {\rm I})$ is the
probability of the data given I, which is usually taken as a
normalising constant; $P({\rm D} | {\rm H, I})$ is the direct probability (or likelihood)
of obtaining the data given H and I. \citet{Berger1987} obtained,
using Bayes' theorem, $p({\rm H}_{0} | \tilde{y})$ with respect to
$p(\tilde{y} | {\rm H}_0)$ and $p(\tilde{y} | {\rm H}_1)$, where ${\rm
H}_{1}$ is the alternative hypothesis.
 \begin{equation}
 p({\rm H}_{0} | \tilde{y}) =\frac{ p({\rm H}_{0}) p( \tilde{y} | {\rm
 H}_{0})}{p({\rm H}_{0})
 p( \tilde{y} | {\rm H}_{0})+p({\rm H}_{1}) p( \tilde{y} | {\rm H}_{1})}.
 \end{equation}
I set $p_0=p({\rm H}_{0})$, and since we have $p({\rm H}_{1})=1-p_0$,
they finally obtained:
 \begin{equation}
 p({\rm H}_{0} | \tilde{y})= \left(1+\frac{(1-p_0)}{p_0}\cal{L}\right)^{-1},
 \label{bayes}
\end{equation}
with $\cal{L}$ being the likelihood ratio defined as:
 \begin{equation}
 {\cal L}=\frac{p(\tilde{y} | {\rm H}_1)}{p(\tilde{y} | {\rm H}_0)}.
 \end{equation}
Here, $p({\rm H}_{0} | \tilde{y})$ is the so-called posterior
probability of ${\rm H}_{0}$ given the observed data $\tilde{y}$.
Naturally there is no way to favour ${\rm H}_0$ over ${\rm H}_1$,
or vice versa, otherwise our own prejudice would most likely be confirmed by the test, i.e. $p_0=0.5$.  Subsequently,
\citet{Berger1997} recommended to report the following when performing
hypothesis testing:
 \begin{equation}
 {\rm if\,\,} {\cal L} > 1, {\rm \,\,reject\,\,} {\rm H}_0 {\rm \,\,and\,\,report\,\,} p({\rm H}_0 | \tilde{y})=\frac{1}{1+ {\cal L}},
 \label{prescrit1}
 \end{equation}
 \begin{equation}
 {\rm if\,\,} {\cal L} \leq 1, {\rm \,\,accept\,\,} {\rm H}_0 {\rm \,\,and\,\,report\,\,} p({\rm H}_1 | \tilde{y})=\frac{1}{1+ {\cal L}^{-1}}.
 \label{prescrit2}
 \end{equation}
The advantage of such a presentation is that even for a borderline
case, say when the ratios above are close to unity, it is clear that
there is only a 50\,\% chance that the H$_0$ hypothesis is wrongly
accepted, or wrongly rejected.  This presentation is more honest and
better encapsulates human judgement and prejudice.

\begin{figure*}
\center{
 \hbox{ \includegraphics[width=0.35\textwidth,angle=90]{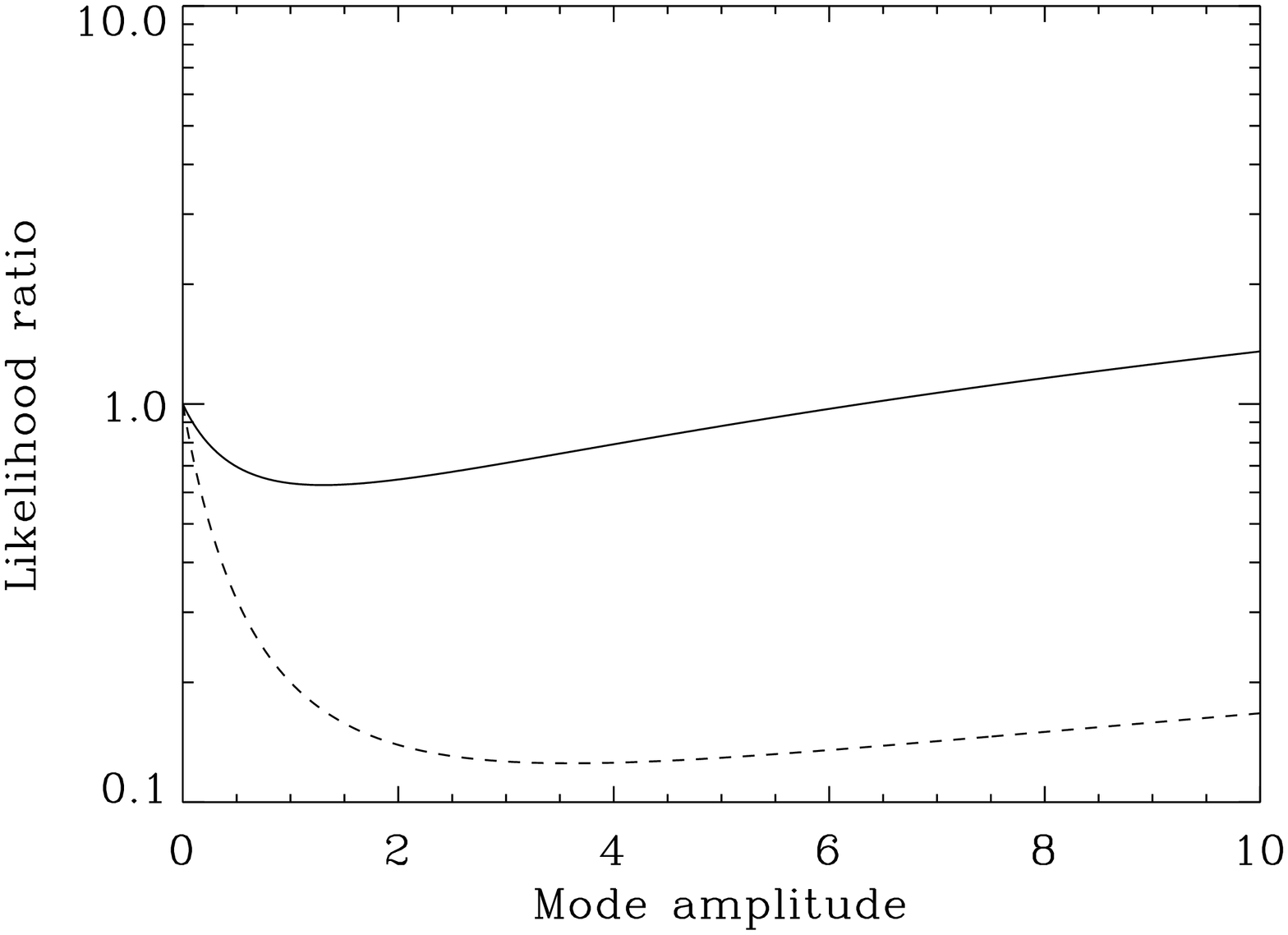}
\vspace{0.0truecm}
\includegraphics[width=0.35\textwidth,angle=90]{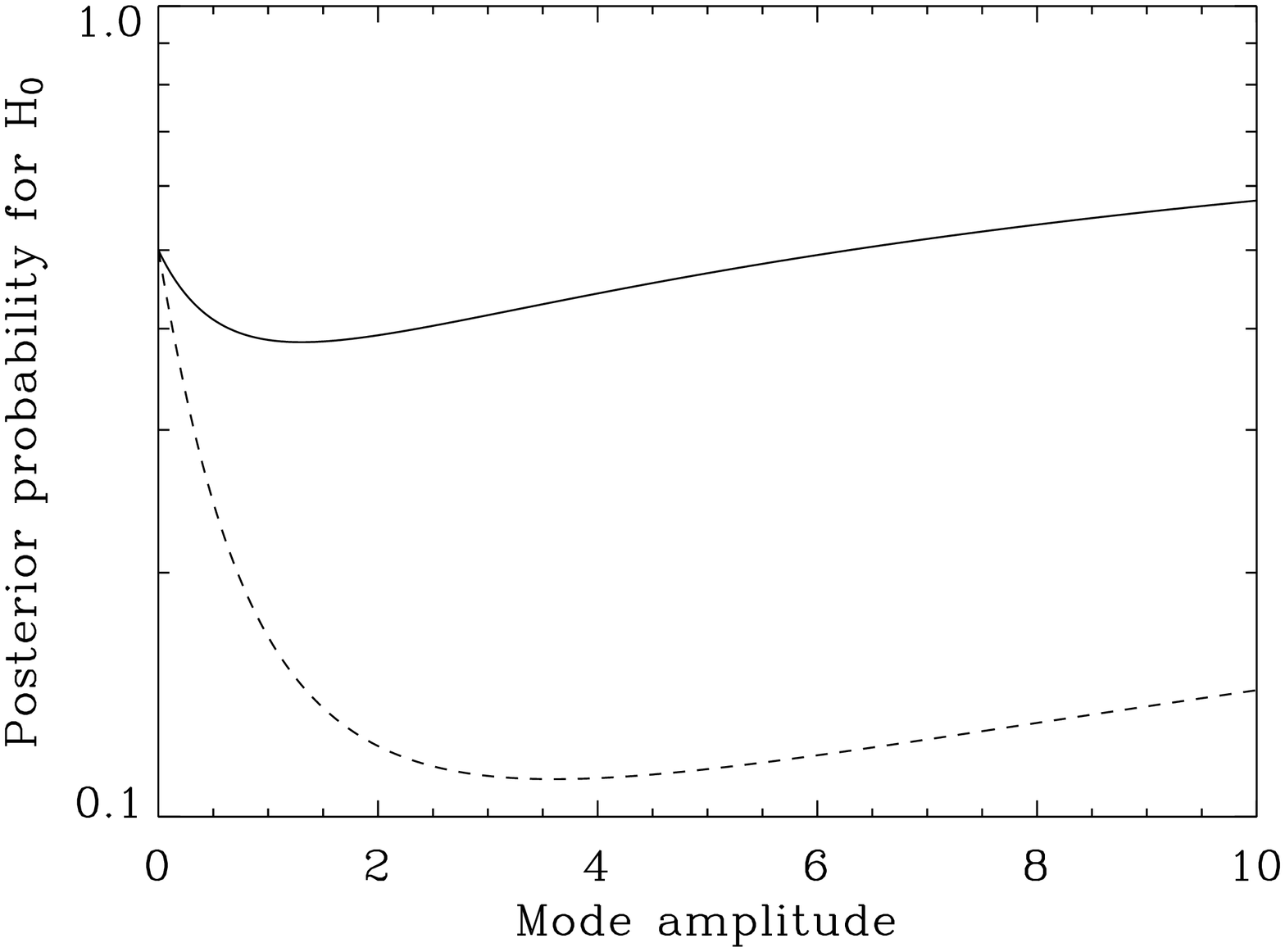}}
}

 \caption[]{On the left-hand side, likelihood ratio ${\cal L}$ as a
 function of the mode amplitude for {\it detection significances} of
 10\,\% (solid line), and of 1\,\% (dashed line); the noise is set to
 unity.  On the right-hand side, the posterior probability of ${\rm
 H}_0$ as a function of mode amplitude for {\it detection
 significances} of 10\,\% (solid line), and of 1\,\% (dashed line)
 (from Eq.~\ref{posterior})].}

\label{bayes-stat}
\end{figure*}

\noindent{\bf Example of posterior probability.} Using the example given in Section~\ref{sec:hypothesis} for the detection of sine waves, we can derive $p({\rm H}_{0} | \tilde{y})$ using
Eq.~(\ref{likelihood}) as:
 \begin{equation}
 p({\rm H}_{0} | \tilde{y})= \left(1+\frac{1}{1+H}p^{-H/(1+H)}\right)^{-1}
 \label{posterior2}
 \end{equation}
where $p={\rm e}^{-\tilde{y}/B}$ is the {\it detection significance}.  Figure~\ref{bayes-stat} show the results for two different {\it
detection significances}.  When the {\it detection significance} is
10\,\%, the likelihood ratio can be greater than unity for large
values of the mode amplitude, leading to the acceptance of the null
hypothesis.  This is rather paradoxical, i.e., that large mode
amplitude can lead to the rejection of the alternative hypothesis.  To
resolve the paradox we note that the posterior probability of ${\rm
H}_0$ is in any case never lower than 40\%, or the posterior
probability of ${\rm H}_1$ is never higher than 60\%.  This implies
that both hypotheses are equally likely when the {\it detection
significance} is as low as 10\,\%.  In other words, when we set, a
priori, a large mode amplitude and get a low {\it detection
significance}, the alternative hypothesis is as likely as the null
hypothesis.  In other words, the assumption about a large mode amplitude is {\it not supported by the data}.

The main conclusion to be drawn from this calculation is that the {\it
detection significance} should be set much lower than 10\,\% in order
to avoid misinterpretation of the result.  For example, with a {\it
detection significance} of 1\,\%, the posterior probability for H$_0$
can fall to 10\,\% when the signal-to-noise ratio is above unity.
\citet{Sellke2001} showed that the posterior probability can never
be lower than the lower bound:
 \begin{equation}
   p({\rm H}_{0} | x) \geq \left(1-\frac{1}{{\rm e} p \ln p}\right)^{-1}
 \label{bound}
 \end{equation}
The reader may verify for themselves that this lower bound is
effectively reached for Eq.~(\ref{posterior2}).  In the case, when
the amplitude of the mode $A$ is not known, one needs to set, a
priori, the value for the likely range of amplitudes.  In the case of a
uniform prior, the posterior probability $p({\rm H}_{0} | \tilde{x})$
then does reach a minimum that is higher than the lower bound of
Eq.~(\ref{bound}) \citep{Appourchaux2009a}.

In summary, the significance level should not be used for justifying a
detection (or a non-detection).  Instead I recommend using the
prescription of \citet{Berger1997}, as given by Eqs.~(\ref{prescrit1})
and (\ref{prescrit2}) and to specify the alternative hypothesis ${\rm
H}_1$.

\vspace{5 mm}

\noindent{\bf On the choice of the prior probability} One important question
when applying Bayesian statistics is what value should the prior
probability of the hypothesis ${\rm H}_0$, i.e., $p_0$, take? We
define the prior probability as the probability that the H$_0$
hypothesis is correct.  The probability that the alternative H$_1$
hypothesis is correct can then be defined as $p({\rm H}_1)=1-p_0$.  It
is common to set $p_0=0.5$ so as to avoid prejudicing one hypothesis
over the other.  Would we expect the probability that H$_1$ and H$_0$ are true to be the same 
in all instances? Since Bayesian statistics requires a priori knowledge, it
is possible to use our knowledge of physics/astrophysics to tell us
which hypothesis is more likely to be true in a given circumstance.

\subsection{Parameter estimation}
The previous section on {\it hypothesis testing} is really the prerequisite when one wants to assess if there is a signal
sought in the observation.  Unfortunately, knowing that a signal is present does not provide any pertinent information for doing physics or astrophysics.  This is the goal of {\it parameter estimation}.

Parameter estimation is a vast subject in statistics.  Below, I will introduce the estimations that are the most commonly used in astrophysics.
The estimations described hereafter are also related to the frequentist and Bayesian world.  As we will see, these estimations are not
so foreign from each other.   The frequentist estimation can be used when the signal-to-noise ratio is high, while the Bayesian estimation is
more useful (but more time consuming) when the the signal-to-noise ratio is low.

\subsubsection{Maximum Likelihood Estimation}
\label{MLE}
As shown in the {\it Historical Overview} section, \citet{Gauss_mle} introduced the concept of {\it Maximum Likelihood Estimation} or MLE.
The aim of MLE is to find the set of parameters that maximise the likelihood of the observed event, this is a {\it point-like} estimation.  Having observed
a random variable $x$ with a probability distribution $p(x,\vec{\lambda})$, where $\vec{\lambda}$ is a vector of $p$ parameters describing the model behind the random variable $x$, the likelihood $L$ of $N$ observations of $x$ is given by:
\begin{eqnarray}
L(x, N, \vec{\lambda})=\prod_{k=1}^{N}  f(x_k,\vec{\lambda}).
\end{eqnarray}
where the product implicitly expresses the fact that all $x_k$ are {\it independent} of each other.  Usually we define the logarithmic likelihood function $\ell$ as
\begin{eqnarray}
\ell (x, N, \vec{\lambda})=\ln{L(x, \vec{\lambda})}= -\sum_{k=1}^{N} \ln f(x_k,\vec{\lambda}).
\label{mle}
\end{eqnarray}
The estimate of $\vec{\lambda}$ is derived from the maximisation of the likelihood as given by Eq.~(\ref{mle}) such that we have
\begin{equation}
\vec{\tilde{\lambda}}=\max_{\vec{\lambda}} \ell(x, N, \vec{\lambda})
\end{equation}
Such an estimator has several interesting properties in the limit of very large sample ($N\rightarrow\infty$) which are:
\begin{itemize}
\item MLE are asymptotically unbiased,
\item MLE are of minimum variance,
\item MLE are asymptotically normal.
\end{itemize}
\noindent The first property implies that:
\begin{eqnarray}
\lim_{N\rightarrow\infty} E(\tilde{\vec{\lambda}}) &=& \vec{\lambda_0}.
\label{large}
\end{eqnarray}
where $ \vec{\lambda_0}$ is the true value of the model.  The second property implies that no other asymptotically unbiased estimator has lower variance.  Using the Cramer-Rao theorem \citep{Cramer1946,Rao1945}, this can be rewritten as:
\begin{equation}
{\rm cov}\left[{\vec{\tilde{\lambda}}}\right] \ge \frac{1}{I(\vec{\lambda_0})}
\label{CramerRao1}
\end{equation}
where $I(\vec{\lambda})$ is the Fisher information matrix whose elements are given by:
\begin{equation}
I_{ij} ={\rm E}\left[\frac{\partial^2{\ell(x,\vec{\lambda}})}{\partial\lambda_i{\partial\lambda_j}} \right] 
\label{CramerRao2}
\end{equation}
\noindent Asymptotically we also have:
\begin{equation}
\lim_{N\rightarrow\infty} {\rm cov}\left[{\vec{\tilde{\lambda}}}\right] =  \frac{1}{I(\vec{\lambda_0})}
\end{equation}
\noindent Finally the third property, regarding the estimator being asymptotically normal, can be expressed as:
\begin{equation}
p(\vec{\lambda})={\cal N}\left(\vec{\lambda_0},\frac{1}{I}\right)
\label{dist_I}
\end{equation}
where ${\cal N}$ is the normal distribution.  This latter equation is usually used for providing the statistical distribution
of \vec{\tilde{\lambda}} as:
\begin{equation}
p(\vec{\lambda}) \approx {\cal N}\left( \vec{\tilde{\lambda}},\frac{1}{H(\vec{\tilde{\lambda}})}\right)
\label{posterior_MLE}
\end{equation}
where $H$ is the so-called Hessian matrix whose elements are derived from:
\begin{equation}
H_{ij} =\left[\frac{\partial^2{\ell(x,\vec{\tilde{\lambda}}})}{\partial\lambda_i{\partial\lambda_j}} \right] 
\label{formal}
\end{equation}
with the property given by the Cramer-Rao theorem as:
\begin{equation}
\frac{1}{H(\vec{\tilde{\lambda}})} \ge \frac{1}{I(\vec{\lambda_0})}
\end{equation}
Equation (\ref {formal}) is used when computing the so-called formal error bars on 
$\tilde{\vec{\lambda}}$; as a matter of fact according to the Cramer-Rao theorem, 
Eq. (\ref {formal}) gives only a lower bound to the error bars.

\noindent{\bf Significance of estimates.} When one uses Least Squares for fitting data, one can test the 
significance of its fitted parameters using the so-called $R$ test \citep{Frieden}.
For MLE, a useful test can be used: the likelihood ratio test.
This method requires maximising first the likelihood ${\rm e}^{-\ell(\omega_{p})}$ of a given event 
where $p$ parameters are used to described the statistical model of the event.  Then if one wants to 
describe the same event with $n$ additional parameters, the likelihood 
${\rm e}^{-\ell(\Omega_{p+n})}$ will be maximised.  The likelihood ratio test consists in making the ratio 
of the two likelihood.  Using the logarithmic likelihood, we can define the ratio $\Lambda$ as:
\begin{equation}
\ln(\Lambda)=\ell(\Omega_{p+n})-\ell(\omega_{p})
\label{ll}
\end{equation}
If $\Lambda$ is close to 1, it means that there is no improvement in the maximised 
likelihood and that the additional parameters are not significant.  On the other 
hand, if $\Lambda \ll 1$, it means that $\ell(\Omega_{p+n}) \ll \ell(\omega_{p})$ and that 
the additional parameters are very significant.  In order to define a significance for the $n$ additional parameters, we need to know the statistics of $\ln(\Lambda)$ under the null hypothesis, i.e. when the $n$ additional parameters are not needed to describe the model.  For this null hypothesis, \citet{Wilks} showed that for large sample size the distribution of $-2$ln$\Lambda$ tends to the $\chi^{2}(n)$ distribution.  

\noindent{\bf Calibration of error bars.}  Before applying MLE to real data, it is always advisable 
to test the power of this approach on synthetic data, i.e. performing 
Monte-Carlo simulations.  They are not merely for playing games; these 
simulations are real tools for understanding what we fit and how we fit it.
Assuming that the statistics of the data is known, performing Monte-Carlo 
is useful for the following reasons:
\begin{itemize}
\item Assessing the model of the data 
\item Assessing the statistical distribution of the parameters of the model
\item Assessing the precision on the fitted parameters of the models
\end{itemize}
The parameters derived by the MLE should have the desirable 
properties of having a normal distribution (See above); if not we advise to apply 
a change of variable on the fitted parameters ($\log x$ for instance).  A normal distribution is necessary to derive meaningful error bars, 
this is the assumption behind Eq. (\ref{formal}).  In order to be able to derive a good estimate of the error 
bars using one realisation, the standard deviation of a large sample of fitted 
parameters should be equal to the mean of formal errors return by the 
fit, i.e. this is the approximation of Eq.~(\ref{dist_I}) by Eq.~(\ref{posterior_MLE}).  In other words we should have the following approximation for the inverse of the covariance
of the parameters:
\begin{equation}
{H(\vec{\tilde{\lambda}})} \approx {I(\vec{\lambda_0})}
\end{equation}
where $H$ is related to the formal error bars and $I$ is related to the asymptotic error bars.  This calibration
as expressed in this equation is key to derive meaningful error bars.  I advise the reader to use such a calibration procedure for checking the formal error bars derived from software codes fitting function using Least Squares.  
The formal error bars derived by a single noise realisation are only {\it a lower limit} to the {\it real} error bars.  This lower limit is never reached when for instance the signal-to-noise ratio is too low.  In this latter case, we are very far from the asymptotic behaviour.  This case is covered by the Bayesian approach to parameter estimation.

\subsubsection{Bayesian parameter estimation}
Parameter estimation can also be done using Bayes' theorem.  In this case, this is the so-called {\it Bayesian inference}.  Using the same notation as before, I can express using Bayes' theorem the probability distribution as:
\begin{equation}
p(\vec{\lambda} | x,{\rm I}) = \frac{p(\vec{\lambda} | {\rm I})p(x | \vec{\lambda}, {\rm I})}{p(x | {\rm I})}
\label{posterior}
\end{equation}
where $\vec{\lambda}$ are the observables for which I seek the {\it posterior probability}, $x$ is the observed data set, and {\rm I} is the information.  The {\it prior probability} of the observables is given by $p(\vec{\lambda} | {\rm I})$: this is the way to quantify our {\it belief} about what I seek.  The likelihood is given by $p(x | \vec{\lambda}, {\rm I})$ which is exactly the $L(x,\vec{\lambda})$ of the previous section.  Therefore the {\it frequentist} approach is related to the {\it Bayesian} approach simply by the {\it frequentist} likelihood, the {\it prior} probability and the normalization factor $p(x | {\rm I})$.  The main advantage of the Bayesian approach is that the posterior probability $p(\vec{\lambda} | x,{\rm I})$ is directly accessible while for the frequentist approach only the location of the maximum of the likelihood is known.  In this latter approach, there is no direct visibility of the parameter probability distribution but only an approximation provided by Eq.~(\ref{posterior_MLE}).  This is why the frequentist approach is a point-like estimation whereas the Bayesian approach is more global.  For instance, the power of the Bayesian approach is such that it provides the full posterior probability which may not be necessarily a normal distribution but could be the sum of many normal distribution due to many local minima.  In that case, only the posterior probability can provide a correct assessment of the statistics of the derived parameters.

\noindent{\bf Posterior probability estimation}  The main difficulty in Bayesian inference is to derive the {\it posterior} probability.  If the derivation of Eq.~(\ref{posterior}) is analytical then parameter estimation can easily be done (See an example in the Application to Asteroseismology section).  When this is not possible, the easiest is to compute the posterior probability by using a random walk algorithm that will provide the posterior probability from a representative samples of the $\vec{\lambda}$.  A famous example of such a procedure is derived from the so-called Metropolis-Hastings algorithm (MH) \citep{Metropolis,Hastings}.  Let us see how this algorithm works in practice for a probability distribution $p(\vec{\lambda})$ for which we want to have a representative sample.  We start from a given point $\vec{\lambda}^{(t)}$ and from a known probability distribution $Q$.  We draw at random from the probability distribution a value $\vec{\lambda}'$ knowing $\vec{\lambda}^{(t)}$ and we compute the following ratio:
\begin{equation}
r=\frac{p(\vec{\lambda}')Q(\vec{\lambda}^{(t)} | \vec{\lambda}')}{p(\vec{\lambda}^{(t)})Q(\vec{\lambda}' | \vec{\lambda}^{(t)})}
\end{equation}
then the new proposed value $\vec{\lambda}'$ is accepted or rejected following this scheme:
\begin{eqnarray}
{\rm If} & r \ge 1 & {\rm then} \,\ \vec{\lambda}^{(t+1)}=\vec{\lambda}' \nonumber \\
{\rm If} & r < 1 & {\rm then}\nonumber \\
	&\vec{\lambda}^{(t+1)}=\vec{\lambda}'  & {\rm if}\,\ r < \alpha  \nonumber \\
	&\vec{\lambda}^{(t+1)}=\vec{\lambda}^{(t)} & {\rm if}\,\ r \ge \alpha
\end{eqnarray}
where $\alpha$ is a random number drawn from a uniform distribution.
Asymptotically the $\vec{\lambda}^{(t)}$ will then tend to have the probability distribution $p(\vec{\lambda})$.  The benefit of this algorithm is that the computation of the normalisation factor in the denominator of Eq.~(\ref{posterior}) is not needed.  Another obvious benefit is that very complex probability distributions can be derived, thereby providing the potential correlations between the various parameters $\lambda_i$.  

The difficulty in the use of the MH is not in the algorithm itself, which is quite easy to implement but in the proper choice of the input distribution $Q$.  This is a vast subject which goes far beyond this course.  The reader will find in \citet{Gregory2005} what is required for delving into the subject of obtaining the posterior probability distribution using various techniques: Gibbs sampling, thermal annealing, convergence and so forth.

\noindent{\bf Mean, rms and moment estimation.} As soon as the {\it posterior probability} is known, we can derive an estimate of the moment $k$ of the parameter $\lambda_i$ 
by deriving first the posterior probability distribution of $\lambda_i$ only or $p(\lambda_i | {x,{\rm I}})$.  This is done by integrating (or marginalising) over the so-called  {\it nuisance} parameters as follows:
\begin{equation}
p(\lambda_i  | x,{\rm I}) = \int_{\Omega_i} p(\vec{\lambda} | x,{\rm I}) {\rm d}\lambda_n
\end{equation}
where $\lambda_{n} \in \Omega_i$ with $n \ne i$.  Then we can compute the moments of the posterior probability by writing:
\begin{equation}
<\lambda_i^k> = \int_{\lambda_i} \lambda_{i}^k p(\lambda_i | {x,{\rm I}}) {\rm d}\lambda_i
\label{moments}
\end{equation}
The first and second moments provide the mean value and rms deviations of $\lambda_i$ as
\begin{equation}
\tilde{\lambda_i}=<\lambda_i^1>
\end{equation}
\begin{equation}
\sigma_{\lambda_i}=\sqrt{<\lambda_i^2>-\tilde{\lambda_i}^2}
\end{equation}
The use of these two moments is enough to describe a normal distribution, as all the other moments can be derived from these two.  Here I note that the Cramer-Rao criteria is also relevant.  It means that the rms value derived above is bounded as follows:
\begin{equation}
\sigma^{2}_{\lambda_i} \geq H^{-1}(\tilde{\lambda}_i)
\end{equation}
As a result, the error bars derived from the Bayesian approach are larger than those returned using MLE. It means that, contrary to popular belief, the Bayesian approach is more conservative than MLE.  

If the posterior probability is not normal, higher moments could be quoted.   It is rather impractical to quote all the moments higher than 2.  Instead, we can use the value of the median and of the percentiles.  We can define, for a random variable $x$ with a probability distribution $p$, two values $x_1$ and $x_2$ such that for a given percentile $q$ we have:
\begin{equation}
q=\int_{-\infty}^{x_1} p(x) {\rm d}x=\int_{x_2}^{+\infty} p(x) {\rm d}x
\label{percentile}
\end{equation}
When $q=50\%$, we have $x_1=x_2$ thereby providing the median.  Usually for a gaussian distribution, the 1-$\sigma$ and 2-$\sigma$ values provide a percentile of 15.9\% and 2.3\%, respectively, i.e. 68.2\% and 95.4\% of the values are in the range defined by $x_1$ and $x_2$.  The use of the 4 percentiles and the median can be shown in a so-called box plot.  The main advantage of using percentiles is that they are invariant under a change of variable.  In other words, when making a change of variable from $x$ to $g(x)$ in Eq.~(\ref{percentile}), the $x_1$ and $x_2$ returned are unaffected by the transform  (provided that $g$ is a monotonic function of $x$).  

In practice, when the analytical integration cannot be done, Eq.~(\ref{moments}) is computed using the MH algorithm mentioned above, such that we have:
\begin{equation}
<\lambda_i^k>= \frac{1}{N_t} \sum_{t} \left(\lambda_{i}^{(t)}\right)^k 
\end{equation}
where $N_t$ is the total number of samples computed returned by the MH algorithm.  Then the median and the percentiles are computed by sorting the values of $\lambda_{i}^{(t)}$.  Examples of such results can be found in \citet{Benomar2009}.

\noindent{\bf Role of the prior.}  The {\it prior probability} expresses what we believe we know (or not) about the parameters $\lambda_{i}$.  The choice of the prior is related to the amount of information at our disposal.  The most obvious {\it prior probability} of the parameters $\lambda_i$ of interest is the one that is uniformly distributed over some range; this is an {\it uninformative} prior.  The role of the prior and its impact on the {\it posterior probability} should ideally be as small as possible.   Objective uninformative priors are derived using the procedure described by \citet{Jeffreys1946}, related to the calculation of the determinant of the Fisher matrix \citep{Fisher_info}. An informative prior could be, for example, a gaussian distribution of a given parameter $\lambda_{i}$.  Priors are not always {\it proper} in the sense that they are not always related to a {\it proper} probability distribution, i.e. providing finite moments.  The $1/\sigma$ prior for the unknown rms value of a parameter is a specific example of an improper prior.  An extensive discussion on the impact of the prior in a Bayesian framework has been well developed by \citet{Jaynes}.

\noindent{\bf Significance and model comparison.}  The power of the Bayesian approach is also to be able to compare different models.  The approach used by frequentists using the likelihood ratio test outlined in the previous section is quite similar to what is called the Bayesian odd ratio.  Let us assume that one wants to compare between different model M$_{n}$.  The odd ratios between any two sets of models is:
\begin{equation}
{\rm O}_{n / m}= \frac{p({\rm M}_n | x ,I)}{p({\rm M}_m | x ,I)}=\frac{p({\rm M}_{n} | I)}{p({\rm M}_{m} | I)} \frac{p(x | {\rm M}_n,I)}{p(x | {\rm M}_m,I)}
\end{equation}
The second part of the equation was derived from Bayes' theorem.  The second fraction closely resembles the likelihood ratio presented above, but it is the product of the ratio of the {\it prior} model probabilities by the ratio of the {\it global} likelihood.  It is termed {\it global} because these probabilities are not a point-like estimate (as in the frequentist approach) but an integration over all the possible values of the estimated parameters $\lambda_i$.  The global likelihood $p(x | {\rm M}_m,I)$ is written as:
\begin{equation}
p(x | {\rm M}_m,I)=\int_{\Omega} p(\vec{\lambda} | {\rm M}_m, I) p(x |\vec{\lambda},{\rm M}_m, I) {\rm d}\vec{\lambda}
\label{global}
\end{equation}
where $p(\vec{\lambda} | {\rm M}_m, I)$ is the prior probability, and $p(x |\vec{\lambda},{\rm M}_m, I)$ is the likelihood.  The computation of the global likelihood is rather difficult but can be done using the MH algorithm under parallel tempering.  The integration of Eq.~(\ref{global}) can then be done using the so-called thermodynamic integration \citep{GelmanMeng}.  Applications of this kind of integration can be found in \citet{Gregory2005}.  

It is also useful to express the posterior probability of each model $p({\rm M}_n | x ,I)$ as follows:
\begin{equation}
p({\rm M}_n | x ,I)=\frac{p({\rm M}_{n} | I)p(x | {\rm M}_n,I)}{\sum p({\rm M}_{m} | I)p(x | {\rm M}_m,I)}
\end{equation}
Usually if the model comparison is an {\it objective} Bayesian analysis, then all prior model probabilities are equal ($p(x | {\rm M}_m,I)=p(x | {\rm M}_k,I), \forall \,\, m,k$).  This assumption can be used when the models are strictly different and are not {\it nested}.  Here {\it nested} means that a child model relies on a parent model,
the former having more parameters describing the model that the latter.  Most of the models we used are indeed {\it nested}, they differ from each other by a few parameters.  In this case, it results in the so-called model {\it multiplicity} that must be taken into account under the {\it subjective} Bayesian approach.  Under this approach, it is possible to have model probability differing from each other ($p({\rm M}_m | I) \ne p({\rm M}_k | I)$).  A by-product of the {\it subjective} Bayesian approach is also to provide an estimate of these $p({\rm M}_m | I)$ based on the data.   Model multiplicity has just been started to be taken into account in model comparison \citep{ScottBerger}.  Since this is very recent, I advise the reader to inform themselves on whether this approach can be useful for model comparison.

\section{Application to asteroseismology}
In the two previous sections, I laid down the foundations for applying harmonic analysis and statistics to astrophysics.  In particular, the field of asteroseismology is extremely relevant for these applications.  Stars have been known to oscillate since, at least, the 16$^{\rm th}$ century when David Fabricius found that $o$ Ceti (Mira) was variable.  Since then many others stars such as $\beta$ Cephei, $\delta$ Scuti, Cepheids, $\gamma$ Dor or solar-like stars have been found to oscillate \cite[See][and references therein]{JCD2004}.
Stellar oscillations are mainly excited by an opacity-driven mechanism ($\kappa$ mechanism) and by turbulence occurring in convection zones \citep{GautschySaio1,GautschySaio2}.  Broadly speaking, the world of stellar oscillations can be divided in two categories:
\begin{itemize}
\item Periodic pulsations having a weakly time-dependent amplitude 
\item Oscillatory eigenmodes stochastically excited
\end{itemize}
The first type results from overstable oscillations in stars being driven by the $\kappa$ mechanism, whose amplitudes are limited by a non-linear mechanism.  The functional form of the variation of the luminosity with time is then periodic but not necessarily sinusoidal. In that case the amplitude of the oscillations varies more slowly that the periods of the oscillations.  

The second type results from modes being randomly excited in solar-like stars whose amplitudes are damped by various mechanisms \citep{GH99}.  In that case, the amplitude of the oscillations varies on time scales shorter than the periods of oscillations.  There are stars for which the two types of oscillations co-exists \citep{Belkacem2009b,Belkacem2010}.

When applying various tools for obtaining the frequencies of the oscillation (frequencies which describe the internal structure of the star), then one should ask oneself which type of functional form the stellar oscillation will have.  Typically there are three type of functional forms:
\begin{itemize}
\item Periodic non-sinusoidal 
\item Sinusoidal
\item Harmonic oscillator stochastically excited
\end{itemize}
These functional forms being periodic the Fourier transform is the obvious choice for time series analysis.  As for the last functional form, the random nature of the excitation imposes the application of a proper statistical treatment of the time series and its associated power spectrum.  Hereafter, I will treat the two most common cases encountered in asteroseismology:  Classical pulsators (periodic non-sinusoidal function), Solar-like oscillators (harmonic oscillators)

\subsection{Classical pulsators}
For stars having luminosity variations whose functional forms is sinusoidal, the common practice is to use the Fourier transform described in the previous sections.  The first application of statistics and Fourier transform was for finding periodicities in earthquakes due to \citet{Schuster}, which has since then be coined the {\it Schuster periodogram}.  

For variable stars (or classical pulsators), the first application of the periodogram is attributed to \citet{Wehlau1964}.  Since that date, the analysis of time series has been evolving to take into account various aspects related to the presence of gaps and the large dynamic range in the amplitude of the periodicities.  One of the problem encountered when observing stars from the ground is that the periodic gaps in the observation (due to the day-night cycle) introduce aliases.  The presence of these gaps produce then spurious peaks located on either side of the main peak located at multiple of $\pm 11.57 \mu$ Hz (1/24 hr).  One solution for taking into account such a frequency response is to apply the CLEAN algorithm \citep{Roberts1987} which is used in radioastronomy for aperture synthesis \citep{Hogbom1974}.  Another approach is to use a combinaison of the CLEAN algorithm and of prewhitening which consists in removing the signal of largest amplitude in the time series (after having being detected) and then to re-compute the periodogram; the procedure iterates until there is no large signal detected in the periodogram \citep{Belmonte1991}.  This latter technique is now the most commonly used for classical pulsators.

\subsubsection{Spectral analysis revisited.} 
\noindent {\bf Single sine wave.}  I already touched upon the Fourier analysis of pure sine waves using the frequentist framework (application of Least Squares).  Here I shall briefly revisit what can be done when using a Bayesian approach.  \citet{Bretthorst}, using a Bayesian approach to the analysis of the time series of a pure sine wave sampled regularly and embedded in noise having a gaussian distribution, demonstrated that the {\it posterior probability} for the frequency $\nu$ of the sine wave can be written as:
\begin{equation}
\ln P(\nu | x, \sigma, {\rm I}) = \frac{C(\nu)}{\sigma^2}  + ...
\end{equation}
where $x$ are the data ($x_i$ taken at time $t_i$), $\sigma$ is the rms value of the noise assumed to be known, and $C(\nu)$ is the Schuster periodogram given by:
\begin{equation}
C(\nu)=\frac{1}{N} \left|\sum_{i=1}^{N} d_i {\rm e}^{{\rm i}2\pi \nu t_i} \right|^2
\label{rms}
\end{equation}
This is nothing less than the power spectrum.  For a pure sine wave with frequency $\nu_0$, the periodogram has a maximum at that frequency.  Using the posterior probability for $\nu$, we then have the first moment of the frequency as:
\begin{equation}
\tilde{\nu}=\int P(\nu | x, \sigma, {\rm I}) {\rm d}\nu=\nu_0
\end{equation}
Following this approach, \citet{Jaynes} demonstrated using Eq.~(\ref{rms}) that for a pure sine wave of amplitude $A$,  the rms error on the frequency $\nu_0$ is given by:
\begin{equation}
\delta \nu = \frac{\sqrt{6}}{\pi} \frac{\sigma}{A}\frac{\sqrt{\Delta t}}{T^{\frac{3}{2}}}
\label{sine_wave_resol}
\end{equation}
where $T$ is the observing time and $\Delta t$ is the sampling time.  This formula is the same as given by \citet{Cuypers1987} and derived by \citet{Koen1999}.  The frequency precision is inversely proportional to the signal-to-noise ratio computed in the time domain.  This equation also shows that the Rayleigh criterion ($1/T$) for frequency resolution is very pessimistic compared to the precision with which frequencies can be measured.  

\noindent{\bf Many sine waves.}  The approach outlined above linking the posterior probability to the Fourier transform justifies {\it a posteriori} the use of that transform: "{\it The highest peak in the discrete Fourier transform is an optimal frequency estimator for a data set which contains a single harmonic frequency in the presence of Gaussian white noise}", \citet{Bretthorst}.  Fortunately, for several harmonic signals whose frequencies ($\nu_i$) are far from each other, the {\it naive} use of Fourier can be broken down in the application of the preceding section as many times as required or:
\begin{equation}
\ln P(\nu_1,...\nu_r | x, \sigma, {\rm I}) = {\left({\sum_{j=1}^{r}\frac{C(\nu_j)}{\sigma^2}}\right)}  + ...
\end{equation}
This justifies the use of the periodogram for finding harmonic signal in a time series  \citep{Bretthorst}.  

\noindent{\bf Periodic signals.} When the signal is periodic but not sinusoidal, the decomposition using the periodogram will provide the fundamental $\nu_f$ of the period and spurious frequencies at $n\nu_f$.  The presence of these spurious frequencies can become difficult to handle when there are a lot of frequencies such as in $\delta$ Scuti \citep{Poretti2009}.  The periodic signal can also be the result of non-linear saturation which cannot be easily described in terms of harmonic signals \citep{Yoachim2009}.   As mentioned above, prewhitening together with the use of the CLEAN algorithm is a possible solution for analysing such periodic signals.  The analysis can also be performed using other techniques such as Principal Components Analysis \citep{Tanvir2005}. 

I showed in a previous section how one could analyse data, that are not equally sampled in time, using the Lomb-Scargle periodogram.  The LS periodogram has been extended not only to a decaying sinusoid \citep{Bretthorst2001a} but also to periodic functions in general \citep{Bretthorst2001b}.  This revisitation of spectral analysis by Bretthorst is extremely useful when one wants to understand how to apply a Bayesian analysis to time series.  Based on the work of \citet{Bretthorst2001b}, a possible application would be to model the amplitude limitation $\delta$ Scuti to obtain the functional forms of the periodic signal.  A different functional form from a sinusoid could be used as an advantage for disentangling the various possible combinations of frequencies \citep{Poretti2009} either due to the gaps, the periodic signal or the physics of the stars.  This is an avenue yet to be explored.

\subsection{Solar-like oscillations}
The analysis of solar-like oscillations is slightly more complicated because of the random excitation of the modes.  In order to apply the Fourier transform and use statistical tools for extracting mode frequencies and mode parameters for such stars, one has to understand how the functional form of the mode amplitude is related to the harmonic oscillator being randomly excited.

\subsubsection{Randomly excited harmonic oscillations}
It is well known that pressure modes (or p modes) are stochastically excited oscillators 
\citep{PK88}.  The source of excitation lies in the many 
granules covering the star \citep[Turbulent convection,][]{GH99,Samadi00I}.  The modes are assumed to be independently excited \citep{PK88} because the eigenfunction of the modes is primarily radial and nearly independent of degree in the upper convection zone, where the modes are excited.  The eigenmodes are harmonic oscillators being intrinsically damped, and excited through a forcing function $F$.  The differential equation of such a damped harmonic oscillator is written as:
 \begin{equation}
 	\frac{{\rm d}^{2}x_{\rm osc}}{{\rm d}t^{2}}+2\pi\gamma\frac{{\rm d}x_{\rm osc}}{{\rm d}t}+(2\pi)^{2}\nu_{0}^{2}x_{\rm osc}=F(t)
 	\label{a}
 \end{equation} 
where $t$ is the time, $x_{\rm osc}$ is the displacement, $\gamma$ is the damping 
term related to the mode linewidth, 
$\nu_{0}$ is the frequency of the mode and $F(t)$ is the forcing 
function.  Equation~(\ref{a}) is also the expression of an auto-regressive (AR) process which in this case is a stationary process.  From this equation the Fourier transform of $x$ can be 
written as:
 \begin{equation}
 	\tilde{x}_{\rm osc}(\nu)=\frac{\tilde{F}(\nu)}{(2\pi)^{2}(\nu_{0}^{2}-\nu^{2}
 	+i\gamma\nu)}
 	\label{b}
 \end{equation} 
where $\tilde{x}_{\rm osc}(\nu)$ and $\tilde{F}(\nu)$ are the Fourier transform 
of $x_{\rm osc}(t)$ and $F(t)$.  From the large number of granules, it can be derived that the forcing function 
is normally distributed.  Therefore the 2 components (the real and imaginary 
parts) of the Fourier transform of the forcing function are also normally distributed (See Section \ref{FFTstat}).  Therefore, for the harmonic oscillator, each component of $\tilde{x}_{\rm osc}(\nu)$ is normally distributed with a mean of zero, and the same variance.  The variance is related to the spectral density given by:
 \begin{equation}	
 	S_{\rm osc}(\nu)=\frac{S_{F}(\nu)}
 	{(2\pi)^{4}[(\nu_{0}^{2}-\nu^{2})^{2}+\nu^{2}\gamma^{2}]}
 	\label{c1}
 \end{equation} 
The spectral density of $\tilde{x}_{\rm osc}(\nu)$ has then a $\chi^{2}$ with 2 degrees of freedom statistics.  For such statistics, I outline that the variance is the same as the mean.  Equation~(\ref{c1}) is the p-mode profile that is usually approximated by a Lorentzian profile when $\nu  \approx \nu_0$ as:
 \begin{equation}	
 	S_{\rm osc}(\nu) \approx \frac{S_{F}(\nu)}
 	{\gamma^2(2\pi)^{4}\nu_0^2}\frac{1}{1+x^2}
 	\label{c}
 \end{equation}
 with $x=2(\nu-\nu_0)/\gamma$.  An asymmetry effect can also be introduced in the profile of  Eq.~(\ref {c}).  The asymmetry effect was first detected with instruments making an image of the Sun \citep{Duvall1993a}.  The asymmetry is due to the location of the excitation of the modes with respect to the resonant cavity.  There is a direct analogy with a source being placed inside a Fabry-Perot cavity \citep{Duvall1993a}.  The asymmetry was then detected by \citet{Toutain1997} with instruments observing the Sun as a star; the sign of the asymmetry depending upon the observables (intensity of velocity).  The empirical theoretical explanation for the different asymmetry in intensity with respect to velocity was given by \citet{Rakesh}.  It is related to the different correlation between the background and the modes in velocity and in intensity which were studied by \citet{Pepe2001}.  The theoretical framework provided then a simple expression for the expression of the mode profile with asymmetry \citep{Rakesh_b}, given by:
\begin{equation}
         S_{\rm osc}(\nu)= H  \frac{1+2 B x}{1+x^2} + B^2
\end{equation}
where $B$ is the asymmetry effect and $H$ is the mode height.  The effect of the linear slope is to change the sign of asymmetry when the sign of $B$ changes.   When $B$ is positive (negative) there is more power at high (low) frequency.  Therefore, if the asymmetry is not taken into account, the sign of $B$ will affect the true location of the mode frequency differently.  This effect, if not taken into account, may provide large systematic errors when comparing different observables thereby providing potential problems for the inferred model of the star.

\subsection{Random non-harmonic field}
Another source contributing to the observed power spectrum is the noise generated by the star itself.  It is quite customary to have the stellar noise increasing at low frequencies.  As a matter of fact, the noise observed is not limited to stars but more related to an intrinsic behaviour encountered in many physical phenomena.  This so-called 1/$f$ noise is so ubiquitous that it is found in almost any physical measurements and also in stars.   The 1/$f$ noise appears in many electrical applications and other applications \citep{Keshner}.  While white noise is known not to have any memory (autocovariance is the Dirac distribution), the $1/f$ noise possesses some memory.  Mathematically, AR models have also these properties.  Random processes following AR models have the desired properties of having memory, thereby producing $1/f$-like power spectra.  This fact was used by \citet{Harvey85} for deriving the solar background noise spectrum for instruments observing solar radial velocities.  The spectrum is derived from a superposition of four components related to solar activity and to three different types of granulation, having different lifetimes.  In \citet{Harvey85}, the power law used had a -2 slope; it was later refined to be arbitrary $b$ \citep{Harvey93} such that we have for the spectral density of background noise $x_n(t)$:
 \begin{equation}	
 	S_{\rm n}(\nu) = \sum_i \frac{H_i}{1+(2 \pi \tau_i \nu)^{b_i}}
 	\label{harvey}
 \end{equation}
where $\tau_i$ is the characteristic time of the decaying autocorrelation function of the process, $H_i$ is the height in the power spectrum at $\nu=0$, and $b_i$ is the power law exponent \citep[with the $b_i$ ranging typically from 2 to 6  for intensity measurements;][]{TA2002,Aigrain2004}.  It is clear that Eq.~(\ref{harvey}) also has a Lorentzian-like shape as much as the harmonic oscillator of the previous section.  Therefore I also outline that modes stochastically excited since they also follow an AR model, have also the same mathematical property as the random non-harmonic field.  

\subsection{Power spectrum model}
\noindent{\bf Stationary processes.}  I showed in Section~\ref{FFTstat} that stationary processes have indeed a $\chi^2$ with 2 d.o.f statistics.  The question is: what is the mean value when different stationary processes are in play?
Let us assume that two time series for two different processes co-exist: $x_{\rm osc}(t)$ related to the excitation of the harmonic oscillators (the eigenmodes), $x_{\rm n}(t)$ related to the non-harmonic noise (the background stellar noise).  The total time series is then given by:
\begin{equation}
x(t)=x_{\rm osc}(t)+x_{\rm n}(t)
\end{equation}
and the autocorrelation is given by:
\begin{equation}
	{\rm E}[x(t_1)x(t_2)]=C_{\rm osc}(\tau)+C_{\rm n}(\tau)+[{\rm E}[x_{\rm osc}(t_1)x_{\rm n}(t_2)]+{\rm E}[x_{\rm n}(t_1)x_{\rm osc}(t_2)]]
	\label{stationary}
\end{equation}
with $\tau=t_2-t_1$ and where $C_{\rm osc}$, $C_{\rm n}$ are the autocorrelation for the harmonic and non-harmonic signals.  The last term in brackets is related to the potential correlation between
the two type of stationary processes.  Usually, we assume that there is no correlation between these processes such that the term in brackets is zero.  Using the Wiener-Khinchine theorem, we have for the spectral density:
\begin{equation}
	S_{x}(\nu)=S_{\rm osc}(\nu)+S_{\rm n}(\nu)
\end{equation}
So the spectral density of the sum of two independent stationary process is the sum of the spectral density of each stationary process.  This implicit assumption, not frequently mentioned, is usually a good approximation for solar-like oscillations.  If we assume that the processes are correlated but that their correlation is stationary we can then write:
\begin{equation}
	S_{x}(\nu)=S_{\rm osc}(\nu)+S_{\rm n}(\nu)+\tilde{I}(\nu)
\end{equation}
where $\tilde{I}(\nu)$ is simply the Fourier transform of the expression in brackets of Eq.~(\ref{stationary}).  Such correlations between the background and the harmonic oscillators have been studied by \citet{Pepe2001}.  The term $\tilde{I}$ is responsible for the mode profile asymmetry as explained above  \citep{Rakesh_b}. 

\noindent{\bf Model of the spectrum.}  The complete model of the power spectrum is given by the superposition of {\it all} the harmonic signals and of the non-harmonic signals.  Apart from the Sun, so far we have only observed disk-integrated oscillations in stars.  Instruments integrating over the stellar surface observe the velocity or the intensity signal as a superposition of various modes of different degrees.  The impact of this integration is to filter out the higher degree modes, keeping typically only the degrees below 4.  The resulting mode sensitivity ($V_l$) was first computed  for velocity measurements for non-rotating stars by \citet{C-DG82}, then by \citet{JCD1989} taking into account Doppler imaging due to rotation;  for intensity, it was first computed by \citet{Toutain1993}.  The impact of the inclination angle of the star upon the mode sensitivity ($c_{l,m}$) was also computed by \citet{Toutain1993}, calculation which was later re-discovered by \citet{Gizon2003}.   The mode sensitivity $V_l$ has also been extensively computed by different authors; e.g.\citet{Bedding1996}, \citet{TA2000} and \citet{Ballot2006}.  

Using the equations given above and taking into account the geometrical mode sensitivity, the full spectral density is therefore given by:
\begin{equation}
S_{x}(\nu)= \sum_{n,l,m}  H_{n}V_l^2c_{l,m}\tilde{L}\left(\frac{\nu-\nu_{nlm}}{\Gamma_{nlm}}\right)+ \sum_i \frac{H_i}{1+(2 \pi \tau_i \nu)^{b_i}}
\label{model}
\end{equation}
where $H_n$ is the mode height for radial order $n$, $\tilde{L}$ is the mode profile, $\Gamma_{nlm}$ is the inverse mode lifetime, $\nu_{nlm}$ is the mode frequency; and the $H_i$, $\tau_i$ and $b_i$ represent the stellar and instrumental background noise (including also the photon noise).  The mode profile is here assumed to be Lorentzian.  It can be replaced by an asymmetrical profile but at the expense of making an incorrect approximation of the correlation effect $\tilde{I}$.  If one wants to properly take into account this correlation effect, one should refer to  \citet{Pepe2001}.  

Several effects will lift the degeneracy of the mode frequency $\nu_{nl}$ such as stellar rotation or stellar activity, thereby splitting the frequency of the mode in different component (Zeeman-like effect).  In this case, the mode frequency can be decomposed into the Clebsch-Gordan coefficient as:
\begin{equation}
\nu_{nlm}=\nu_{nl}+\sum_{i=1}^{i_{{\rm max}}} a_i (n,l) l {\cal P}_{l}^{(i)}(m/l) 
\end{equation}
where $i_{\rm max} \le 2l$, and $a_i(n,l)$ are the splitting coefficients and ${\cal P}_{l}^{(i)}$ are polynomials given by  \citet{RITZ91} or by \citet{Chaplin2004}.  The advantage of the ${\cal P}_{l}^{(i)}$ is that they are by construction orthogonal to each other.  With this property, the max $i$ for the decomposition can be different from $2l$, in other words removing or adding polynomials will not affect the value of the $a_i(n,l)$.  To second order, this property may not be completely true as there could be slight variations in the mode height due to stochastic excitation or other effects not modelled.  For example, the observation of the integrated signal over the stellar surface with a star observed perpendicular to its rotation axis ($i=90^{{\rm o}}$) will {\it remove} modes for which $l+m$ is odd.  If this averaging effect is not taken into account the decomposition will affect the value of the $a_i(n,l)$ \citep{Chaplin2004}.

Depending on the model used, different assumptions concerning the mode and background parameters are possible.  For instance, the mode linewidth can be assumed to be independent of frequency, or to vary with frequency according to some polynomials, or to be independent for each order $n$, or to be independent for each degree $l$, and so forth and so on.  It is sometimes desirable to reduce the number of fitted parameters using such assumptions.  Examples of such assumptions are found in the references given in the next two sections.

\subsection{Frequentist parameter estimation}
Now having observed a star,  we have a time series of intensity or velocity fluctuations which is properly sampled, and for which the filtering effect of the data reduction is understood.  We compute the Fourier transform and then obtain the power spectral density properly scaled using Parseval's theorem.  The amount of gaps is negligible such that the frequency bins are independent of each other.  The statistics of the power spectrum is $\chi^2$ with 2 d.o.f.  Assuming that all the stationary processes are nearly independent, we have a complete model of the power spectrum including harmonic and non-harmonic signals.

Now that I have set the scene, all the actors are in place for the play.  The method applied for deriving all the parameters of Eq.~(\ref{model}) is usually to use MLE.  The use of MLE for fitting solar power spectrum was first mentioned by \citet{TDJH86}, then applied and tested by \citet{EA90}; in which they also mentioned the use of Monte-Carlo simulations for validating the method.  Their pioneering work led in helioseismology to the understanding of the derivation of error bar for mode frequencies \citep{KL92,TTTA94} which is given by:
\begin{equation}
\delta \nu=\sqrt{\frac{\Gamma}{4\pi T}} (\beta+1)^{\frac{1}{4}}\left(\sqrt{\beta+1}+\sqrt{\beta}\right)^{\frac{3}{2}}
\end{equation}
This equation shows that the mode precision is proportional to the square root of the mode linewidth and is proportional to the square root of the frequency resolution.  This results greatly contrasts with that of a pure sine wave given by Eq.~(\ref{sine_wave_resol}).  Similarly other error bars were also derived for linewidth, mode height, white noise \citep{TTTA94}; rotational splitting \citep{TTTA94,Gizon2003,Ballot2008}; inclination angle \citep{Gizon2003,Ballot2008}; and of the correlation between the mode height, the linewidth and background noise \citep{TTTA94} (See Appendix A); and between the rotational splitting and the inclination angle \citep{Ballot2008}.  

Monte-Carlo simulations have also been used for verifying the veracity of using the inverse of the Hessian for estimating error bars on the parameters of the model  \citep{EA90,TTTA94, JSTB94, TALG98, Gizon2003} (See Section~\ref{MLE}).  When doing so, it was patent that the distribution for mode height, linewidth and background noise was not normal but log-normal.  There are two main reasons for this state of affairs: first, these simulations are far from the asymptotic behaviour, otherwise the distribution would be normal; second, the mathematical transform $\log(x)$ allows us to have a distribution closer to a normal distribution than $x$ alone.  In other words, the asymptotic behaviour is reached faster with such a transform.  The same would apply to any transform reducing the amount of value larger than the median with respect to value smaller than the median\footnote{The reader may convince themselves by simulating a $\chi^2$ with n d.o.f for example together with a $\log x$ or $\sqrt{x}$ transform.}.

Although the MLE has been in use for more than 20 years by helioseismologists, it is not yet completely adopted by our fellow asteroseismologists.  The body of evidence provided by  helioseismology is rather large, hereafter I will necessarily refer to a few examples: IPHIR (InterPlanetary Helioseismology by IRradiance measurements) photometers aboard the Russian mission Mars94 \citep[Sun as a star;][]{TTCF92}, VIRGO (Variability of solar IRradiance and Gravity Oscillations) photometers aboard SoHO (Solar and Heliospheric Observatory) \citep[Sun as a star and low resolution;][]{CF97,TABA97}, GOLF (Global Oscillations at Low Frequency) spectrometer aboard SOHO \citep[Sun as a star,][]{Gelly2002}, BiSON (Birmingham Solar Oscillations Network) instrument \citep[Sun as a star][]{WJC1996}, LOWL (Low-$l$) instrument \citep[imager;][]{JS92,JSTB94}, GONG (Global Oscillation Network Group) instrument \citep[imager;][]{FH96,TALG98}, SOI/MDI (Solar Oscillations Investigation / Michelson Doppler Imager) instrument  aboard SoHO \citep[imager;][]{AK97}.  

Following these successful applications for solar data, MLE was preliminarily tested on synthetic stellar timeseries for use on the CNES CoRoT mission \citep{Appourchaux2007,Appourchaux2007b}.  The recipe laid down in this latter publication was found later not be applicable to the real CoRoT data.  \citet{Appourchaux2008} introduced a scheme using not fit on narrow frequency window \citep[{\it local} fit as in][]{Appourchaux2007} but using fit on large frequency window ({\it global} fit).  This new scheme has now been applied to solar-like stars and red giants observed by CoRoT \citep{Garcia2009,Barban2009,Mosser2009, Deheuvels2010,Baudin2011}; and to solar-like stars observed by the NASA (National and Aeronautics Space Administration) Kepler mission \citep{Mathur2011}.

One of the recent controversies in asteroseismology is related to the degree identification in the power spectra of HD49333 (observed by CoRoT).   In \citet{Appourchaux2008} the degree of the modes could not easily identified leaving two possible solutions: the ridges in the echelle diagram were either $l=0-1$ or $l=1-0$.  The reason for such a difficult identification, already anticipated by \citet{Appourchaux2007}, is related to a mode linewidth larger than in the Sun combined with a narrower small spacing than in the Sun.  As a result, the two ridges $l=0-2$ and $l=1$ were indistinguishable from each other.  One of the solutions found by \citet{Appourchaux2008} was to base the choice of the ridge upon the likelihood ratio test.  Here I would like to cite a warning in \citet{Appourchaux2008}  as to the use of this test:
"{\it However, it is important to remember that a higher likelihood does not mean that a given model is physically more meaningful, rather, it means the model is statistically more likely.}"  It was clear at this stage that in order to take into account prejudices and a priori knowledge that could have helped to get the physically meaningful model, a Bayesian approach to the problem was required (ibid).  This is the subject of the next section.

\subsection{Bayesian parameter estimation}
In asteroseismology, the first attempt at applying a Bayesian approach was due to \citet{Brewer2007}.  They used a model based upon pure sine waves which did not correctly model stochastic excitation of harmonic oscillations.  Their model can be used for classical pulsators or when the modes have a lifetime longer than the observation time.  Following the attempt by \citet{Brewer2007}, I realised that the approach and framework provided by \citet{Bretthorst} is not generally applicable in asteroseismology.  The separation in time between the deterministic signal and the noise cannot be done when oscillators are stochastically excited.  Since the separation is not possible, the statistics of the stellar signal suggested by \citet{Brewer2007} needs to be replaced by the likelihood usually used by frequentist \citet{Appourchaux2008b}.  At the time, it was not clear that the replacement was so {\it obvious}.

Following the early work of \citet{Appourchaux2008b}, the application of Bayesian analysis to solar-like oscillators has been quite extensively developed.  The difficult data analysis of HD49333 led to the development of several Bayesian data analyses in asteroseismology  \citep{Benomar2009,Gruberbauer2009}.  A simplified Bayesian approach called Maximum A Posteriori (MAP) has also been used for putting constraints on the mode height \citep{Gaulme2009}.  Recently, \citet{RHTC2011} produced a very nice guide for anyone wanting to do Bayesian data analysis in asteroseismology.  A more extensive guide in French is also available in \citet{Benomar2010}.  These two guides covers many aspect of the application of Bayes' theorem such as power spectrum models, statistics, priors, parallel tempering and global likelihood.  Here I should add that the Bayesian framework has been used not only for parameter estimation but also for making decision based upon posterior probability estimates: mode detection \citep{Appourchaux2009a,AMB2010}, presence of $l=3$ modes and mixed modes \citep{Deheuvels2010}.

As anticipated by \citet{Appourchaux2008}, the HD49333 controversy led to the first useful application of the Bayesian approach.  \citet{Benomar2009} using the same time series as \citet{Appourchaux2008} applied a Bayesian analysis and derived the global likelihood of the various models previously used.  In doing so, \citet{Benomar2009} showed that amongst the four models used the identification of \citet{Appourchaux2008} was the most probable one {\it but} for a model having {\it no} $l=2$ modes !  When these latter are included \citep[as in][]{Appourchaux2008}, \citet{Benomar2009} found that the identification of \citet{Appourchaux2008} was less probable (14\%) than the other identification (23\%); thereby showing that both models were nearly equiprobable.  The absence of $l=2$ modes in HD49333 was also the conclusion of \citet{Gruberbauer2009} who also used a Bayesian approach.  The main differences of the work of \citet{Gruberbauer2009} from that of \citet{Benomar2009} are that they do not use global likelihoods for decision and that their model has no splitting (no degree identification possible).  The controversy was finally closed when 180 days of data were available for HD49333.  Using these data, \citet{Benomar2009b} showed that both frequentist and Bayesian approaches favour a model having $l=2$ modes with a degree identification different from that of \citet{Appourchaux2008}.  

What lessons can we draw from such a controversy?  The process of the advancement of science cannot be done without making {\it errors}.  The can be of several kind: 
\begin{itemize}
\item Done on purpose for hiding the facts (a lie)
\item	Done by ignorance of the facts
\item Done in knowledge of the facts
\end{itemize}
The first kind of error has lead to several infamous publication such as a cold fusion \citep{Fusion} finding which was never reproduced.  The second kind of error is due to lack of information or knowledge; in this world of information flow this can easily happen to anyone of us.  The third kind can either be perceived as an error from an outsider or as a decision done in full conscience of what has been decided.  The {\it error} of \citet{Appourchaux2008} is obviously akin to the last kind.  If we would have had the choice, we would have certainly opted for applying the Bayesian approach in \citet{Appourchaux2008b} but time prevented us to do so.  This is another lesson to be learnt: even in this fast paced world of information flow and publications, one should never trade quality of publication and scientific integrity for fast publication.  Science is like pasta, a slow sugar.

\section{Conclusion}
If you have been able to survive until now, I would like to thank you.  It would be pretentious to believe that this {\it Conclusion} is going to conclude anything for good.  As a matter of fact, this conclusion aims at giving the reader the incentive to go beyond this {\it course} but still using this {\it course} as a foundation for the future.  Data analysis clearly starts from the instrumentation itself, this aspect was not treated here.  It is clear that the measurement leading to the analysis of time series is done through an instrument.  From this point of view, the aspects related to the filtering due to the instrument are implicitly laid out.  Apart from these mere details, the reader should find all that is needed for a proper understanding of data analysis in asteroseismology.  Needless to say, that this field is evolving very rapidly, thanks to the CoRoT and Kepler missions.  In the last couple of years, the Bayesian data analysis has developed quickly.  Despite what is commonly thought, the Bayesian approach although using prejudices is in the end more conservative than the frequentist approach.   Actually, the main drawback of this approach is not in the methodology itself but the speed at which the computation of the results can be performed.  

With the advent of the new PLATO mission, it can be anticipated that both Bayesian and frequentist data analysis will be used for deriving the mode parameters of more than 20,000 stars.  I can think of a hybrid scheme where both approaches will be used depending on the signal-to-noise ratio in the power spectrum.  Another aspect which was partly discussed here is how one can assess and insure that the fitted mode parameters are of usable quality for doing stellar models.  This is certainly to be one of the future challenges of asteroseismology: the automatic assessment of stellar mode parameters.  I can also anticipate that we will apply quality assessment which are regularly used in factories doing mass production.  For instance, sampling at random some of our output products may be useful for {\it qualifying} the lot as useable by the scientific community.  We are really in the infant stage for such massive quality assessment.  

The future of asteroseismology will reach another milestones when the surface of stars will be imaged.  At that point time, we will then be able to probe the internal and dynamics structure of stars in the very same manner as we did with the Sun.  The Stellar Imager could be the next asteroseismic challenge whose achievement may be not so far in the future \citep{jcd2011}.

As for any new endeavour, we need to learn from the past for going ahead while keeping in mind Timothy Leary's motto: {\it Think for yourself, question authority}.

\begin{acknowledgments}
\noindent {\it{\bf Acknowledgments.}  I would like to thank Pere Pall\'e for inviting me to give these lectures in the beautiful island of Tenerife.  I would also like to express my thanks for the many discussions I had with my long standing colleagues from which this course benefited: Fr\'ed\'eric Baudin, Othman Benomar, Patrick Boumier, William Chaplin, Patrick Gaulme, Laurent Gizon and Takashi Sekii.  This course also benefited from the acute reading of John Leibacher.  Last but not least, many thanks to my wife, Maryse, and my sons, K\'evin and Thibault, for their indefectible support; without forgetting the mention of the purring cat, Myrtille.  Thanks to the volcano of El Teide for momentary time support, going up there was not an easy ride !}
\end{acknowledgments}

\appendix
\section{Correlation between mode height and linewidth}
\citet{TTTA94} showed that some fitted mode parameter were intrinsically correlated.  When one wants to derive the mode amplitude $A$ (proportional to the square root of $\Gamma H$) one should take into account the correlation between the errors on $\Gamma$ and $H$ which are the mode linewidth the mode height, respectively.  Since we have $A \propto \Gamma H$, and using the logarithm of these quantities, we have:
\begin{equation}
\log A= \frac{1}{2} (\log \Gamma + \log H) + ...
\end{equation}
or simply
\begin{equation}
a= \frac{1}{2} (\gamma + h) + ...
\end{equation}
The error bar on $a$ is derived from:
\begin{equation}
\sigma_a^2=\frac{1}{4}\left(\sigma_{\gamma}^2+\sigma_{h}^2+2{\rm E}(ah)\right)
\end{equation}
which can be rewritten using the correlation coefficient as:
\begin{equation}
\sigma_a^2=\frac{1}{4}\left(\sigma_{\gamma}^2+\sigma_{h}^2+2\rho \sigma_{\gamma}\sigma_{h}\right)
\end{equation}
Using the work of \citet{TTTA94}, and after some mathematical manipulation, I can derive for a single mode the correlation as:
\begin{equation}
\rho=\frac{{\rm E}(\gamma h)}{\sigma_h \sigma_\gamma}=-\sqrt{\frac{\sqrt{\beta+1}}{\sqrt{\beta+1}+\sqrt{\beta}}}
\end{equation}
where $\beta$ is the noise to the mode height ratio in the power spectrum.  This correlation is negative, and its absolute value always greater than 76.5\% when $\beta < 1$.
It means that even when the modes are easily detected the correlation is always very large and never negligible.  Finally the error bar on $a$ is then given by:
\begin{equation}
\sigma_{a}=\frac{1}{2 \sqrt{ T \Gamma \pi}} \sqrt{\left(\sqrt{\beta+1}+\sqrt{\beta}\right)\left(\left(\sqrt{\beta+1}+\sqrt{\beta}\right)^3-4(1+\beta)\sqrt{\beta}\right)}
\end{equation}
where $T$ is the observation time.  From this equation, I can deduce that the precision on the measurement of the mode amplitude increases with the observation time and the mode linewidth.

% \clearpage

\bibliographystyle{aa}
\bibliography{thierrya}
\end{document}